\documentclass[11pt]{article} 
\usepackage[utf8]{inputenc}
\usepackage{amsmath,amssymb,amsfonts} 
\usepackage[UKenglish]{babel} 
\usepackage[T1]{fontenc}
\usepackage{setspace}
\usepackage{xcolor}
\usepackage{graphicx}
\usepackage{lmodern}
\usepackage{booktabs}
\usepackage{ae,aecompl}
\usepackage{fullpage}
\usepackage[toc,page]{appendix}
\usepackage{epstopdf}
\usepackage{import}
\usepackage{caption}
\usepackage{subcaption}

\usepackage{todonotes}
\newcommand{\dd}{ \mathcal D}
\newcommand{\ns}{\sigma}
\newcommand{\corner}{\mathcal J}
\newcommand{\tcorner}{\mathrm{T}\corner}
\newcommand{\tk}{j}
\newcommand{\hs}{\hat s}
\newcommand{\ts}{\tilde s}

\newcommand{\cg}{\mathbf g} 
\newcommand{\ch}{\mathbf h} 
\newcommand{\red}[1]{{\color{black}#1}}

\allowdisplaybreaks 

\graphicspath{{figures/}}

\newcommand{\comp}{\mathbb C}

\newcommand{\cM}{\mathcal M}
\newcommand{\cN}{\Sigma }
\newcommand{\cJ}{\mathcal J}

\newcommand{\dn}{{\nabla}}

\newcommand{\dM}{\partial \cM} 

\newcommand{\action}{I_{\mathrm{EH}}}

\newcommand{\tpsig} {\mathrm{T_p}\,\cN}
\newcommand{\tsig}{\mathrm{T}\, \cN}
\newcommand{\ei}{e_{\scalebox{0.6}{(i)}}} 
\newcommand{\ej}{e_{\scalebox{0.6} {(j)}}} 
\newcommand{\eij}{e_{\scalebox{0.6}{(i,j)}}} 
\newcommand{\nij}{n_{\scalebox{0.6}{(i,j)}}} 
\newcommand{\Lij}{{\Lambda_{\scalebox{0.6}{(ij)}}}} 

\def\be{\begin{equation}}
\def\ee{\end{equation}}

\begin{document}
\title{Boundary and Corner Terms \\in the \\Action for General Relativity} 
\author{Ian Jubb${}^\ast$, Joseph Samuel${}^\dagger$, Rafael Sorkin${}^\ddagger$ and Sumati
  Surya${}^\dagger$ \\ 
${}^\ast$\textit{\small{Blackett Laboratory, Imperial College, London, SW7 2AZ, UK}}, \\ 
${}^\dagger$\textit{\small{Raman Research Institute, C.V. Raman Avenue, Sadashivanagar, Bangalore 560 080, India}}\\
${}^\ddagger$\textit{\small{Perimeter Institute, Waterloo, Canada}}}

\maketitle

\begin{abstract}We revisit the action principle for general relativity, motivated by the path integral approach to
  quantum gravity.  We consider a spacetime region whose boundary has piecewise $C^2$ components,
  each of which can be spacelike, timelike or null and consider metric variations in which only the
  pullback of the metric to the boundary is held fixed. Allowing all such metric variations 
  we present a unified treatment of the spacelike, timelike and null boundary components using 
  Cartan's tetrad formalism. Apart from its
  computational simplicity, this formalism gives us a simple way of identifying corner terms.  
  We also discuss ``creases'' which occur when the boundary is the event horizon of a black hole.
  Our treatment
  is geometric and intrinsic and we present our results both in the computationally simpler tetrad
  formalism as well as the more familiar metric formalism. We recover known results from a simpler 
  and more general point of view and find some new ones.
\end{abstract} 

\section{Introduction}

The Einstein-Hilbert (EH) action for general relativity depends on the metric and its first and second
derivatives. Indeed, the dependence on second derivatives is forced on us by the principle of
general covariance since, there is no local coordinate scalar that can be formed from the metric and
its first derivatives. 
By an appropriate choice of coordinates, we can make the first derivatives vanish at any point so that
the only candidate for the action is the cosmological constant term.  

While the EH Lagrangian does depend on the second derivatives of the metric, the dependence is rather
innocuous since, as it turns out, the equations of motion are second order in metric
derivatives, rather than fourth order, as one might naively expect. One can remove the dependence on
second derivatives by adding a total divergence to the EH Lagrangian, which integrates to a boundary
term. The appropriate action for general relativity is therefore the  EH action with this  boundary
term. This makes the action first order in the metric derivatives: the second derivative 
term $\partial \partial g$ present in the Einstein Hilbert Lagrangian is replaced by a term of the form $(\partial g)^2$.
All of this has been known for a while \cite{york,GH}. 

Our first reason for revisiting the action principle for General Relativity is the path integral
approach to quantum gravity. In summing over histories, we would like the quantum amplitudes to have
the ``folding''  property: 
\begin{equation} 
K(X_1,X_3)=\int dX_2 K(X_1,X_2)K(X_2,X_3), 
\label{feynmanone}
\end{equation}  
where $X_1$ and $ X_3$ are initial and final states respectively and $X_2$ is an intermediate state
which is summed over. In the metric representation $X_1, X_3$ represent the metrics on an initial
and final spatial hypersurface $\Sigma_{1,3}$ and $(\Sigma_2, X_2)$, an intermediate spatial
geometry. We would clearly like  the action to be additive 
under a decomposition of spacetime into pieces.
There is a close relation between additivity of the action and having a first order Lagrangian. This can be clearly seen in a particle mechanics
analogy. Consider the amplitude for a particle to go from $x_0$ at time $t_0$ to $x_N$ at time $T=t_N$. $K(x_0,t_0;x_N,T)$. Introducing time slices
at $t_k=k\epsilon=kT/N$, we have the skeletonised version of the path integral
\begin{equation} 
K(x_0,t_0;x_N,T)=\int dx_1dx_2...dx_{N-1}  K(x_0,t_0;x_1,t_1)K(x_1,t_1;x_2t_2)...K(x_{N-1} t_{N-1};x_N,T), 
\label{feynmantwo}
\end{equation}  
If the Lagrangian is first order, i.e. if $L$ depends only on $x$ and ${\dot x}$, the additivity of
the action is immediate. One writes the short time propagator replacing ${\dot x}$ in the Lagrangian
by $(x_{k+1}-x_k)/\epsilon$. This results in nearest neighbour couplings on the time lattice with the
sites labeled by $k$. Decomposing the lattice into two parts separated by $t_j$ gives us the folding
property Eqn (\ref{feynmanone}). However, for a second order Lagrangian $L(x,{\dot
  x},{\ddot x})$, one needs {\it three} time steps in order to define ${\ddot x}$. E.g ${\ddot
  x}_k=(x_{k+1}+x_{k-1}-2 x_k)/\epsilon^2$.  This brings in {\it next} nearest neighbour couplings on
the time lattice, which spoils the additivity of the action.

A related point stems from the tensor nature of the gravitational field, which is not captured in the simple particle analogy above.
In summing over histories that go from $X_1$ to $X_3$ via $X_2$ we allow all spacetime
geometries, which on pullback agree with $X_2$. No further restriction
needs to be placed on the metric. In particular, the components of the metric
in directions transverse to the spacelike surfaces need not be held fixed.  Textbook treatments (see
\cite{wald,poisson} for example) however hold {\it all} components of the metric fixed on the
boundary, which is a stronger requirement. In a path integral, one typically sums over all paths
without requiring continuity of all components of the metric across $\Sigma_2$. 
All we need is that the pullback of the four-metric to $\Sigma_2$ agrees with $X_2$.

Our second reason for revisiting the action principle is to explore boundaries of different
signatures. A region in spacetime may have  boundaries with components which are  spacelike,
timelike and null. There may also be corners where components  of the boundary join. 
We present a formalism in which all these cases are derived in a transparent manner. 
The role of boundaries in gravitational physics has been increasing in recent years. Ideas relating 
bulk and boundary degrees of freedom have been discussed in the context of black hole entropy. One of the possible
applications of our work is in black hole physics. 

{The need for adding a total divergence to the Einstein-Hilbert action was realised very early in the history of General Relativity\cite{einstein}.
The required boundary counterterm was given a geometric interpretation by 
York\cite{york} and this line of thought was
carried further by Gibbons and Hawking in their work on 
black hole thermodynamics. When the boundary has corners, there is a need for
additional corner terms. These were first discussed by Sorkin and Hartle \cite{Sorkinthesis,SorkinHartle}, and
  subsequently by Hayward\cite{Hayward:1993my}, Brill and Hayward\cite{Brill:1994mb} for timelike
  and spacelike boundaries.}  The need for a treatment of null boundaries was recognised by Parattu
et al \cite{Parattu:2015gga,Parattu:2016trq}. There are also several contributions by
Neiman\cite{Neiman:2012fx,Neiman:2013ap,Neiman:2013taa,Neiman:2013lxa} and \red{Epp\cite{Epp:1995uc}}. Very recently Lehner et al
\cite{LMPS} have given a detailed account of this problem. Our work differs from all these in
several respects. \red{We postpone a discussion of the differences to the concluding section.}

Our treatment uses both the tetrad formulation and the metric formulation. We present a unified approach to  
the different boundary signatures.  
Indeed, as will appear below, this
simplifies the calculation considerably. In section
\ref{two} we review some of the mathematical preliminaries. In Section
\ref{three} we present the tetrad formulation, which brings out the need for the corner terms and 
their explicit forms. 
In Section \ref{four}, we perform the whole calculation in the metric formulation, which is more familiar to readers. 
Section \ref{five}
contains a discussion and some open questions.

\section{Mathematical Preliminaries}\label{two} 
The spacetime manifold $({\cM},g_{ab})$ is described by a Lorentzian metric $g_{ab}$, where $a,b$ are spacetime indices going over $(0,1,2,3)$. 
Our signature
is $(-+++)$. We begin with  the Einstein-Hilbert action  
\begin{equation} 
\action = \frac{1}{2} \int d^4x \sqrt{-g} R  
\end{equation} 
for a spacetime $(\cM,g_{ab})$, where $\partial\cM=\cup_i \cN_i$ can have several piecewise $C^2$
components $\cN_i$ whose normals $n_{ia}$ are everywhere either timelike, spacelike or null. We have chosen units in which $8\pi G$ has been set to 1.

\begin{figure}[t!]
  \centering
    {\includegraphics[scale=0.7]{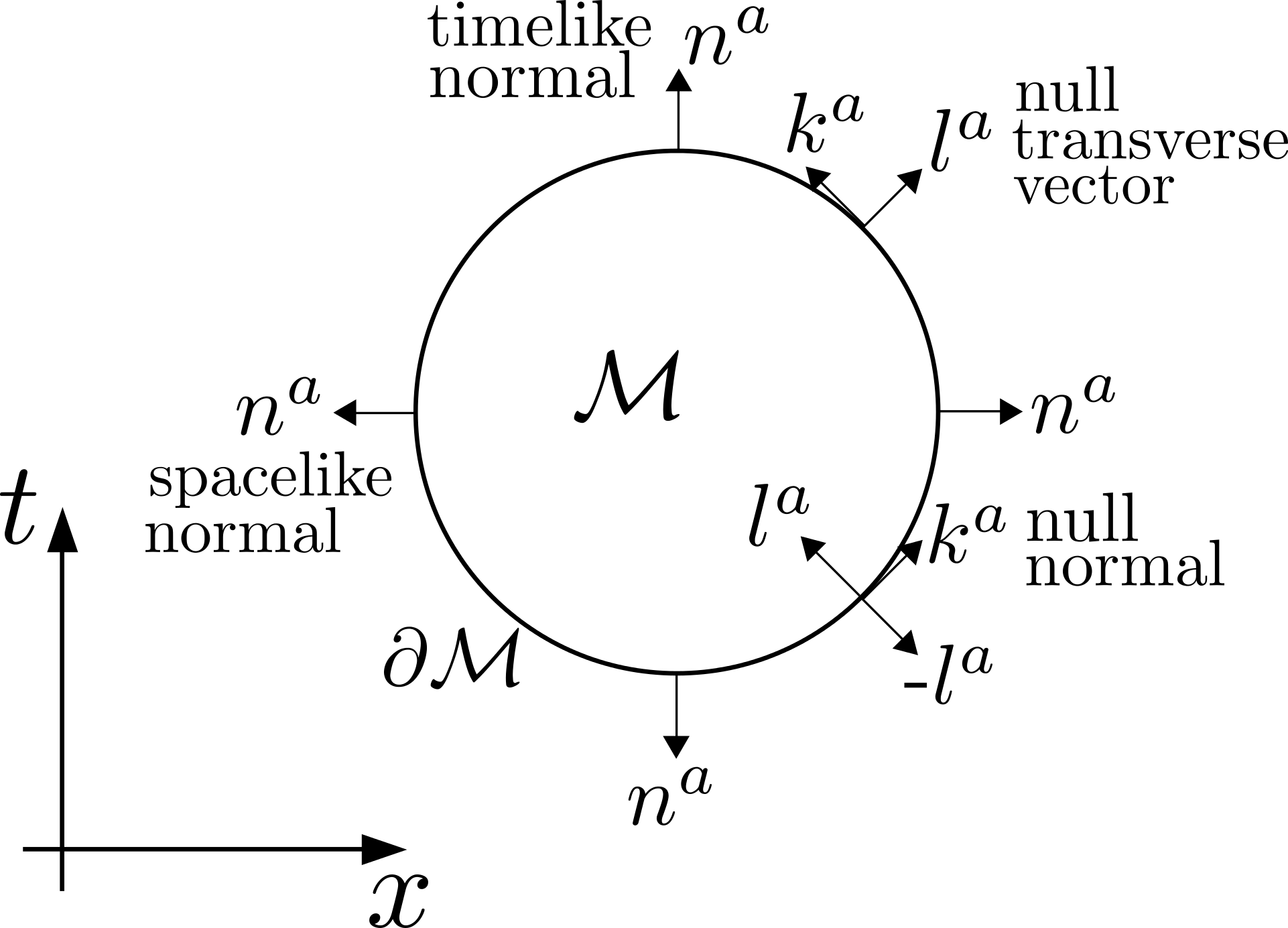}}
     \caption{An illustration of how the normal and transverse vectors would be orientated on a patch of $1+1$ Minkowski spacetime whose boundary is a circle.}
     \label{fig:transverse_and_normal_vectors}
\end{figure}

Consider a single component of the boundary $\cN \subset \partial\mathcal{M}$.  When $\cN$ is non-null, the
unit normal $n_a$ satisfies $n^an_a=\epsilon$ where $\epsilon\equiv \pm 1$ depending on whether
$\cN$ is timelike or spacelike.  When $\cN$ is null the normal $n_a$ is not unique, but for each
$n_a$ there is an equivalence class of null vectors {$l^a$} 
which satisfy $n_al^a=-1$.  In order to unify
the treatment of the null and non-null cases, in addition to the normal $n_a$ 
to $\cN$ we will find
it useful to define a {\sl transverse} vector $Q^a$ to $\cN$ which does not lie in $\tpsig$.  For
non-null $\cN$,  $Q^a$ is proportional to $ n^a$, i.e., the transverse and normal directions
  coincide upto a sign. For null $\cN$ the natural
choice is  $Q^a \propto l^a$. It is this identification of the transverse vector $Q^a$ which helps unify
our treatment, rather than the normal vector $n^a$. For a smooth boundary in spacetime, for example, it is
not the normal that gives a continuous or consistent definition of the outward direction, but rather the
transverse vector, as shown in Figure~\ref{fig:transverse_and_normal_vectors}.
The metric can be decomposed into components along $\cN$ and transverse to it, so that   
\begin{eqnarray} 
\label{hypersurface} 
g_{ab} &=& h_{ab} + \epsilon n_a n_b \quad(\mathrm{Non-Null})  \nonumber \\
g_{ab} &=& \sigma_{ab} - l_an_b - n_al_b \quad(\mathrm{Null}) 
\end{eqnarray} 
where $h_{ab}$ is the induced metric on $\cN$ and $\sigma_{ab}$ is the induced metric on the spatial
slice $\ns \subset \cN$ normal to $n_a$.

The ``joins'' or intersections $\cJ_{ij} = \cN_i \cap \cN_j$ of $\dM$ are allowed to be
discontinuous in the sense that $n_i^a$ and $n_j^a$ differ at $\cJ_{ij}$. The $\cJ_{ij}$ are of
codimension two and, like the boundary components, may also be timelike, spacelike or null.  
By considering the span of the normals $n_i$ and $n_j$ and looking at the range of the polynomial
$f(\alpha)=(n_i+\alpha n_j)^2$ as $\alpha$ varies over the real line, one easily arrives at the
following classification: the plane of the two normals in the tangent space has Lorentzian signature
and the join is spacelike if i) at least one of the normals is timelike, or ii) both the normals are
null, or iii) one normal is spacelike and the other null, with $n_i.n_j\neq0$ or iv) both normals
are spacelike and $n_i.n_j> 1$.  The plane of normals has Riemannian signature and the join is
timelike if i) both normals are spacelike and $n_i.n_j< 1$.  Finally, the plane of normals is null
and the join is null if i) one normal is null and the other space like with $n_i.n_j=0$.

In Section \ref{three}  we use the Cartan tetrad formalism. This has the significant advantage offered by differential forms which
can be integrated over manifolds without reference to a metric or its signature. It also has the
advantage of giving us a fiducial Minkowski vector space as a reference. 
Given a metric $g_{ab}$ on $\cM$ we choose an orthonormal frame such that $g_{ab}=e^\mu_a e^\nu_b
 \eta_{\mu\nu} $.  
The tetrad $e^\mu_a$ maps a vector $X \in T_p \cM$ to a point in $(\cM_0,\eta_{\mu\nu})$ 
\begin{equation} 
e^\mu_a:X \rightarrow e^\mu(X)= e^\mu_a X^a = X^\mu\in \cM_0,
\label{map}
\end{equation}    
where $(\cM_0, \eta_{\mu\nu})$ is a fixed fiducial Minkowski vector space, with $a$ the spacetime
index and $\mu$ the frame index ranging over $0,1,2,3$.  The map Eqn 
(\ref{map}) is invertible, since we
assume that the metric is non-degenerate.  The spacetime metric $g_{ab}$ is then the pullback of the
fiducial metric $\eta_{\mu\nu}$ on $\cM_0$.  Frame indices $\mu,\nu$ are raised and lowered with
$\eta_{\mu\nu}$.  There is an $O(1,3)$ gauge freedom in the choice of the $e_a^\mu$.  Associated
with the $e^\mu_a$ are the connection 1-forms $A^\mu_{a\nu}=e^\mu_b \nabla_a e_\nu^b$ where
$\nabla_a$ is the metric compatible Christoffel connection. 
$A$ takes values in the Lie Algebra of
$O(1,3)$ and is antisymmetric in the frame indices: $A^{\mu \nu}=-A^{\nu \mu}$.
$A$ is compatible with frames and satisfies Cartan's equation
\begin{equation}
de^\mu+A^\mu{}_\nu\wedge e^\nu=0. 
\label{compat}
\end{equation}
Written explicitly in the 
spacetime indices, the field strength of $A$ is 
\begin{equation} 
F_{ab}^{\mu \nu}=\nabla_a A_b^{\mu \nu}-\nabla_b A_a^{\mu\nu} +  A^\mu_{a \rho}  A_b^{\rho \nu} - A^\mu_{b \rho}  A_a^{\rho \nu}=2 \dd_{[a} A_{b]}^{\mu\nu}
\end{equation} 
which is more succinctly expressed as $F^{\mu\nu} = d A^{\mu \nu}+ A^\mu_\rho \wedge A^{\rho
  \nu}=2\dd A^{\mu \nu}$ where the wedge product is with respect to the spacetime indices.

Using the  algebraic identity
\begin{equation}
\tilde{\eta}^{abcd} \epsilon_{\mu\nu\rho\lambda} e^\rho_a e^\lambda_b=-e (2!)^2 \, e_\mu^{[c}e_\nu^{d]}  
\label{identity}
\end{equation} 
where $e=\sqrt{-g}$ and $\tilde{\eta}^{abcd}$ is the Levi-Civita 
tensor density,  the Einstein-Hilbert action\footnote{Note that we do not
  regard this as a first order Palatini action, since $A$ is a function of $e^\mu_a$ determined by
  Eqn (\ref{compat}) and is not
  independent.}   takes the form 
\begin{equation} 
I_{EH} = \frac{1}{4}\int d^4 x \, \epsilon_{\mu \nu \rho \lambda} \, \ e^\mu \wedge e^\nu
\wedge F^{\rho \lambda}. 
\end{equation} 
{The variation of $I_{EH}$ gives us a bulk term (which yields the equations of motion) and a boundary term.
The boundary term will be expressed as the variation of a boundary action $-I_{B}$, which gives us a counterterm to
be added to the action.
The total gravitational action is therefore
\begin{equation}
I_G= I_{EH} + I_B 
\end{equation}    
where $I_B$  in the non-null case is the Gibbons-Hawking-York
(GHY)  term. From its definition, the boundary term $I_B$ is only defined up to terms that have zero variation. Certain
imaginary terms that have been discussed before in the literature are of this variety. We will ignore them for the most part and comment on them
in the conclusion.}  
When the  boundary is only piecewise $C^2$ the
boundary contribution includes ``corner'' terms. It is the evaluation of these various boundary
components in the tetrad formulation that we will now focus on.

\section{The Tetrad formalism}\label{three} 
\subsection{Boundary Terms}
Varying the Action $I_{EH}$ 
we find
\begin{equation} 
\delta I_{EH} = \frac{1}{4} \biggl(2 \int_\cM \epsilon_{\mu \nu \rho
  \lambda} \, \delta e^\mu \wedge e^\nu \wedge F^{\rho \lambda} + 2\int_\cM  \epsilon_{\mu \nu \rho
  \lambda} \, e^\mu \wedge e^\nu \wedge D \delta A^{\rho \lambda} \biggr). 
\end{equation} 
The first term gives us Einstein's vacuum equations in the form 
$e^\nu\wedge F^{\rho \lambda}\epsilon_{\mu\nu\rho\lambda}=0$
and the second term reduces to a  boundary contribution 
\begin{equation}
\label{boundaryterm1} 
-\delta I_{B}= \frac{1}{2} \int_{\cM} D(\epsilon_{\mu \nu \rho \lambda} \, e^\mu \wedge e^\nu \wedge \delta A^{\rho \lambda} )=
+\frac{1}{2} \int_{\partial\mathcal{M}} \epsilon_{\mu \nu \rho \lambda} \, e^\mu \wedge e^\nu \wedge \delta A^{\rho \lambda} 
 \end{equation} 
 In order to calculate the boundary term we require that the pullback of the metric to the boundary
 $\cM$ is unvaried. We can, in addition, demand that the pullback of $e^\mu$ to the boundary
 $\cM$ also has zero variation. This permits us to take $\delta$ outside the integral and express it
 as the variation of
\begin{equation}
\label{boundaryterm2} 
-I_{B}=
\frac{1}{2} \int_{\partial\mathcal{M}} \epsilon_{\mu \nu \rho \lambda} \,  e^\mu \wedge e^\nu \wedge  A^{\rho \lambda}.
\end{equation}

Our derivation so far is independent of the type of boundary $\partial\mathcal{M}$.  We will now
show that this expression is  the GHY  term  written in a universal form, by looking at the three
types of boundaries  -- spacelike, timelike and null.  We now choose {\sl adapted tetrads}  so that one of the 1-form fields $e^\mu$ is  normal
to the boundary.  The natural choices for the spacelike, timelike and null normals to  $\partial\mathcal{M}_{s,t,n}$  are $e_a^0=n_a$, $e_a^1=n_a$ and
$e_a^\pm=n_a$, respectively, where $e_a^\pm=(e_a^0\pm e_a^1)/\sqrt{2}$. In general  we  write $n_a=e_a^\alpha$
where $\alpha=0,1,\pm$ depending on $\partial \mathcal{M}$. 
Let us also always choose $n^a$ to be  outward directed
for  $\partial\mathcal{M}$, a well-defined concept for the non-null normals. For null normals, it is
the transverse null tetrad $l_a$ defined by $l_an^a=-1$ which must be outward pointing. This picks
the orientation of $n_a$.  For example in Minkowski spacetime  if $n_a=e_a^+$, then the transverse null tetrad $l_a=e_a^-$ is
is chosen with the same time  orientation as  $e_a^+$ so that $e_a^+e^{a -}=-1$.  

Using the relation Eqn (\ref{identity}) we see that the integrand in Eqn (\ref{boundaryterm2}) can be
simplified. Using the notation $n_a=e^{\hat{\alpha}}_a$, where the hat indicates that $\alpha$ is a fixed index ($0,1,\pm$), for which the summation convention does not apply,
we have 
\begin{equation} 
  e n_a e_\mu^a e_\nu^b A_b^{\mu \nu}=  e \delta_\mu^{\hat{\alpha}} e_\nu^b A_b^{\mu\nu}= e e^b_\nu A^{\hat{\alpha}\nu}_b
\end{equation}
where the sum over $\nu$ extends over all indices {\it except} $\hat{\alpha}$ because of the antisymmetry of $A$ in the frame indices. Putting in the form of $A$ we have
\begin{equation} 
-e e^b_\nu e^\nu_c \nabla_b e^{\hat{\alpha}c} = - e(\delta^b_c-e_{\hat{\alpha}}^b e^{\hat{\alpha}}_c) \nabla_b n^c, 
\end{equation}
which gives the universal  boundary term 
\begin{equation} 
\label{uboundary} 
I_B= \int_{\partial\mathcal{M}} e(g^{bc}-e_{\hat{\alpha}}^b e^{\hat{\alpha}c}) \nabla_b n_c.  
\end{equation} 
Note that we have made no assumption above regarding extending the normal $n_a$ off the boundary. The normal is only defined at points on the boundary
and we only use its tangential derivatives.

Observe  further that  in the adapted tetrads for non null normals ($\hat{\alpha}=0,1$)
\begin{equation}
e_{\hat{\alpha}}^b e^{\hat{\alpha}c}=\eta_{\hat{\alpha}\hat{\alpha}}e^{\hat{\alpha}b}e^{\hat{\alpha} c}=\epsilon n^b n^c
\end{equation} 
and for null normals ($\hat{\alpha}=+$)
\begin{equation}
e_{\hat{\alpha}}^b e^{\hat{\alpha}c}=\eta_{+-}e^{-b}e^{+c}=-l^b n^c
\end{equation} 
Using the decomposition Eqn (\ref{hypersurface}) we note that 
\begin{eqnarray} 
(g^{bc}-e_{\hat{\alpha}}^b e^{\hat{\alpha}c}) \nabla_b n_c &=& (g^{bc} -\epsilon n^b n^c)\nabla_b n_c= h^{bc}\nabla_b n_c=K \quad(\mathrm{Non-Null})  \\
(g^{bc}-e_{\hat{\alpha}}^b e^{\hat{\alpha}c}) \nabla_b n_c&=& (\sigma^{bc} -n^bl^c)\nabla_b n_c= (\Theta - \kappa)\quad(\mathrm{Null})  \\
\end{eqnarray} 
where $K=h^{ab}\nabla_a n_b $ is the extrinsic curvature of $\partial \cM_{t,s}$, 
$\Theta=\sigma^{ab}\nabla_a n_b$ the null expansion on $\partial\cM_n$ and
the surface gravity $\kappa=l^an^b\nabla_bn_a$ measures the failure of $n^a$ to be affinely parameterised.   
For $\partial\mathcal{M}_{s,t} $ this gives  the expected GHY term 
\begin{equation} 
I_B= \int  \sqrt{\pm h} K\,d^3x,  
\label{nonnull}
\end{equation}
where $x$'s are coordinates on the boundary.
For $\partial\mathcal{M}_{n}$ this gives 
\begin{equation}
I_B= \int  \sqrt{\sigma} (\Theta-\kappa) d\lambda d^2 x,  
\label{null}
\end{equation}  
where the $x$'s are now spatial coordinates on the null surface and $\lambda$ is a parameter along the null generator satisfying $n^a\partial_a=\frac{\partial }{\partial \lambda}$. 

The boundary term Eqn (\ref{boundaryterm2}) is not gauge invariant under $O(1,3)$ transformations
(although its variation is). This is because $A$ transforms inhomogenously by
\begin{equation}
A\rightarrow \Lambda^{-1}A \Lambda +\Lambda^{-1}d \Lambda
\end{equation}
with the result that 
\begin{equation}
\label{gtrans} 
I_B\rightarrow I_B+\frac{1}{2}\int_{\partial\cM} \epsilon_{\mu\nu\rho\lambda} \, e^\mu\wedge e^\nu \wedge \cg^{\rho\lambda} 
\end{equation}
where $\cg=\Lambda^{-1}d\Lambda$ is in the Lie Algebra of $O(1,3)$.

We note that in the adapted tetrads there is a residual gauge freedom in the little group $H$, which
preserves the normal. The little group is given by $H=O(3)$ for timelike, $H=O(1,2)$ for
spacelike and $H=E(2)$ for null normals. It is easily checked that the adapted boundary term {\it
  is} invariant under gauge transformations of the little group. In fact for $\Lambda\in H$, $\ch=\Lambda^{-1}d \Lambda$
satisfies $\ch^{{\hat{\alpha}}\lambda}=\ch^{\rho{\hat{\alpha}}}=0$ for $\hat{\alpha}$ a fixed index labelling the normal, as above.
For vector fields $t^a$, tangent to the boundary, we have $e^{\hat{\alpha}}(t)=n_at^a=0$ 
and so the 
change in $I_B$ under a gauge transformation
\begin{equation}
\label{htrans} 
\Delta I_B=\frac{1}{2}\int_{\partial\cM} \epsilon_{\mu\nu\rho\lambda} \, e^\mu\wedge e^\nu \wedge \ch^{\rho\lambda} 
\end{equation}
vanishes entirely, since the four indices of $\epsilon_{\mu\nu\rho\lambda}$ must all be distinct for a nonvanishing contribution.

We  introduce four discrete elements $D$ of $O(1,3)$ corresponding
to each of the connected components of the group. They
are I, P, T and PT, where P and T stand for parity and 
time-reversal respectively. Since these are constant matrices, 
the connection $A$ transforms homogeneously and the 
boundary term Eqn (\ref{boundaryterm2}) is invariant under such transformations.
These discrete elements $D$ will be needed in the next section to relate frames across a 
join.  

\subsection{Corner Terms}
\label{corner}
The fact that the boundary term Eqn (\ref{boundaryterm2}) is not gauge invariant can be exploited to
identify the corner terms. By adapting our frame to the normal we have been able to derive the forms
Eqn (\ref{nonnull},\ref{null}) of the boundary GHY terms for all signatures of the boundary. When there is a join of two boundary
components, the adapted frames will not, in general, agree at the join. In order to pass from one frame to the other we will use the
following procedure.  By means of a gauge transformation in the little group $H$, we will ensure that two of the frame fields from each boundary component
are tangent to the join and agree with each other at the join. By use of discrete elements in $O(1,3)$, we will ensure that the frames are related by an element in the identity
component of $O(1,3)$. With these choices, the relation between the two frames is a Lorentz transformation in the 2-dimensional plane of normals.
The change in the boundary term Eqn (\ref{boundaryterm2}) under this $O(1,3)$ gauge transformation gives us the corner terms.

Let $\cN_i$ and $\cN_j$ meet along a join $\cJ_{ij}$. The boundary term  Eqn (\ref{uboundary}) is
valid when an adapted frame is used, but the latter changes in going from $\cN_i$ to $\cN_j$.  
This corresponds to effecting an $O(1,3)$ transformation in  Eqn (\ref{boundaryterm2}) which relates the
adapted  frames $\ei^\mu,\ej^\mu$ of  $\cN_i,\cN_j$. By operating on the frames by  discrete elements $D$ of $O(1,3)$, we
can arrange that the two frames are related by an element in the identity component of $O(1,3)$.

For spacelike  joins,  by further gauge transformations in the little group, one can  arrange that $\ei^2=\ej^2$ and $\ei^3=\ej^3$ and that both 
of these are orthogonal to the timelike plane of normals.  The two frames $\ei$ and $\ej$ are
therefore related by a Lorentz boost in the timelike plane of normals, 
\begin{equation}
\ei^\mu={\Lij}^{\mu}_\nu\ej^\nu. 
\end{equation}
We define the discontinuous gauge transformation $\lambda \in O(1,1)$ to be the identity on  $\cN_i$ and 
$\Lij$ on $\cN_j$ 
\begin{equation}
\lambda_{ij}=\exp{[\eta K \Theta^{(H)}_{ij}]},
\end{equation}
where $\Theta^{(H)}_{ij}$ is the Heaviside  function that takes values 
$0$ on $\cN_i$ and 1 on $\cN_j$,  $\eta$ is the rapidity parameter and $K$ 
the boost generator in the plane of
normals. $\cg^{\rho\sigma}=(\Lij^{-1} d\Lij)^{\rho\sigma}=\eta K^{\rho \sigma} d\Theta^{(H)}_{ij}$  is
therefore proportional to a delta function that is peaked on the join  $\cJ_{ij}$ and vanishes on  
$\cN_i$ and $\cN_j$. The gauge transformation of the boundary term Eqn (\ref{gtrans})
results in the join term 
\begin{equation}
\frac{1}{2}\int_{\cJ_{ij}} \eta \epsilon_{\mu\nu\rho\sigma} e^\mu\wedge e^\nu  K^{\rho \sigma},
\end{equation}
which in this case (since only $K^{01}=-K^{10}$ is non vanishing), simplifies to 
\begin{equation}
I_{\cJ_{ij}}= \int_{\cJ_{ij}}e^2\wedge e^3 \eta= \int_{\cJ_{ij}} dA \eta,
\end{equation}
where $dA$ is the area element of the join.

It is possible to express the rapidity that appears in the corner term using the angle between
the normals. The Lorentz boost with rapidity $\eta$ can be written as $\ej^+=(\exp{\eta})
\ei^+,\ej^-=(\exp{-\eta}) \ei^-$ and the  timelike and spacelike normals as  $\nij=(\eij^+\pm
\eij^-)/\sqrt{2}$, respectively. Using the symbols $T,S,N$ to denote a timelike , spacelike or null
normal,  respectively, we find that if the two normals at the join are  (i)   TT: 
$n_i.n_j=-\cosh{\eta}$, (ii)  TS: 
$n_i.n_j=\sinh{\eta}$, (iii)  TN:  $n_i.n_j=-\exp{\eta}/\sqrt{2}$, (iv) SS:  $n_i.n_j=\cosh{\eta}$,
(v) SN:  $n_i.n_j=\exp{\eta}/\sqrt{2}$ and (vi) NN:$n_i.n_j=-\exp{\eta}$.

For timelike joins, the argument is very similar.  We can by gauge transformations in the little
group arrange that $\ei^0=\ej^0$ and $\ei^1=\ej^1$ and that both of these are orthogonal to the
spacelike plane of normals.  The two frames $\ei^\mu$ and $\ej^\mu$ are now related by a rotation in the
spacelike plane of normals
\begin{equation}
\ei^\mu={\Lij}^\mu_\nu  \ej^\nu.
\end{equation}
Again, define the discontinuous gauge
transformation $\lambda\in O(2)$ as  the identity on $\cN_i$ and 
$\Lij$ on $\cN_j$,  so that 
\begin{equation}
\lambda_{ij}=\exp{[\eta J \Theta^{(H)}_{ij}]}, 
\end{equation}
where $\eta$ is now the rotation angle and $J$ the rotation generator in the plane of normals. Again
this gives rise to a contribution from the join
\begin{equation}
\frac{1}{2}\int_{\cJ_{ij}} \eta \epsilon_{\mu\nu\rho\sigma} e^\mu\wedge e^\nu J^{\rho\sigma}.
\end{equation}
Since the nonvanishing components of $J$ are $J^{23}=-J^{32}$, we have the form of the corner term:
\begin{equation}
I_{\cJ_{ij}}= \int_{\cJ_{ij}}e^0\wedge e^1 \eta=\int_{\cJ_{ij}} dA \eta,
\end{equation}
where $dA$ is the area element of the join. 
Relating the inner products to the angles follows the case of spacelike joins and we do not repeat the analysis here.
A salient difference is that the angles are only defined modulo $2\pi$. This arises because the group $SO(1,1)$ is simply connected
($\pi_1(SO(1,1))=0$), while the group 
$SO(2)$ is multiply connected ($\pi_1(SO(2))=\mathbb{Z}$). This ambiguity does not
however  affect the variation.

\red{Null joins differ in that the plane of 
normals and the tangent space to the join share a one dimensional, null subspace. 
If $n_i$ is spacelike and $n_j$ is null (with $n_i.n_j=0$), $n_j$ belongs {\it both} to the span of normals and the tangent space to the join.
It is possible to adapt a null Lorentz frame to both $\cN_i$ and $\cN_j$ as follows: $e_i^+=e_j^+=n_j$, $e^3_i=e^3_j=n_i$ and $e^2_i=e^2_j$, $e^-_i=e^-_j$. Since
$e^\mu_i=e^\mu_j$, we have $\Lij$ equal to the identity and $\eta=0$. The corner term therefore  vanishes.}

\subsection{Creases}
A physically interesting situation covered by the above analysis occurs when one of the boundaries
of spacetime is the event horizon of a dynamically evolving black hole. In this case the horizon does not remain smooth when new generators
enter or leave the horizon. Suppose that we are interested in 
the boundary of a  future set. (The case of past sets is similar). 
The boundary of a future set is ruled by null generators. However, when these null generators cross because of gravitational focussing effects, they leave the boundary
and enter into the interior of the  future set. 
The horizon then develops a caustic, generically a spacetime region of codimension two, where the normal to the wavefront is discontinuous.
When this happens,  we have a ``crease'' which separates regions of the null surface with different normal vectors. Locally, this is
no different from a null-null join discussed above. From the analysis already presented we would expect a boundary
term to appear as an integral along the crease of the rapidity parameter, just as in the NN case treated above.

\section{The metric formalism}
\label{four} 
While the tetrad formulation is calculationally simpler, it is also true that the metric formulation is 
more familiar to most readers. In this section, we present the metric formulation of the above
calculation, which has also recently been given in  \cite{LMPS}. 
To find the boundary contribution to the action we need to consider the most general class of
variations $\delta g^{ab}$ which leave the induced metric on $\cN$ fixed, so that for any $t^a, s^a
\in \tsig$,
\begin{equation} 
 \delta g_{ab} t^a s^b =0.
\end{equation} 
Using $\delta(g_{ab}g^{ac})=0$ to relate the variation of the covariant and
contravariant metrics  we find that 
\begin{equation} 
\label{lowerupper}
\delta g_{ab} = - g_{bd} g_{ac} \delta g^{cd} \,\,  \Rightarrow \,\, \delta g^{ab} t_a s_b =0.
\end{equation} 
From the decomposition Eqn (\ref{hypersurface}) of $g_{ab}$ into components transverse to and along
$\cN$,  we see  that the most general variation takes the form  
\begin{equation}
\label{gpert} 
\delta g^{ab} = 2 n^{(a} \delta Q^{b)},  
\end{equation}  
where we have made the identification 

{\begin{equation}
\label{Qdef}
Q^a=
\left\{
	\begin{array}{ll}
		  \epsilon \, n^a  &  (\mathrm{Non-null}) \\ 
		  - \,  l^a  &  (\mathrm{Null})
	\end{array}
\right.
\end{equation}}
The 4-vector $\delta Q^a$ therefore gives the full  admissible $10-6= 4$ parameter degrees of freedom in this class of
variations. 

It is useful to decompose $\delta Q^a$ into components   transverse to and along $\cN$ 
\begin{equation}
\label{pert} 
\delta Q^a = \alpha Q^a + t^a, 
\end{equation}     
where $t^a \in \tsig$, 
and such that $n^at_a=0$ for both null and non-null cases.  When $\cN$ is
null $t^a$ can be further decomposed as
\begin{equation}
\label{npert}  
t^a=\beta n^a + s^a \quad \mathrm{(Null)}
\end{equation}  
where  $s^an_a=0$.  
 
Using the unperturbed metric to raise and lower
indices, the variation of the covariant quantities is 
\begin{equation} 
\delta Q_a = \delta g_{ab} Q^b + g_{ab} \delta Q^b, \quad \delta n_a = \delta g_{ab} n^b + g_{ab}
\delta n^b, 
\end{equation} 
which along with  Eqns (\ref{lowerupper}-\ref{npert}) simplifies 
to the general expression  
\begin{eqnarray} 
\label{covvar} 
\delta n_a & = & -\alpha n_a,  \\ 
\delta Q_a &=& \beta n_a,  
\end{eqnarray} 
where $\alpha,\beta$ are independent when $\cN$ is null and $\beta=-\epsilon \alpha$ when $\cN$ is non-null. 
The parameter $\alpha$ can moreover be related to a variation of the volume element in
both the null and non-null cases  
\begin{equation}
\alpha=-\delta (\ln  \sqrt{-g}) 
\end{equation}
where we have used  $\delta (\ln  \sqrt{-g}) =-\frac{1}{2} g_{ab} \delta g^{ab}=-\alpha g_{ab}n^a
Q^b$ using 
Equations (\ref{gpert}) and (\ref{pert}). 
The boundary term resulting from the variation of the Einstein-Hilbert action has the general
form 
\begin{equation} 
\label{boundary}
-\delta I_B = \frac{1}{2}\int_{\cN}  dV v^a n_a  
\end{equation} 
where $v^a=-g^{ab} C_{cb}^c+ g^{cb} C_{cb}^a$ and $dV$ is the volume element on $\cN$ and 
$C_{ab}^c$ is the variation in the metric compatible connection  
\begin{equation} 
C_{ab}^c = \frac{1}{2} g^{cd} \{ \dn_a \delta g_{bd} + \dn_b\delta g_{ad} -\dn_d \delta g_{ab}\}. 
\end{equation} 
with  $\dn_a$ the connection compatible with  $g_{ab}$.  

The task is then to find the boundary term $I_B$ which has to be added to the Einstein-Hilbert
action.  The integrand in Eqn (\ref{boundary}) can be simplified to
\begin{equation} 
\label{ehboundary} 
v^an_a = -  n^a g^{bc} \{ \dn_a \delta g_{bc} -\dn_b \delta g_{ac}\}, 
\end{equation} 
for all types of  $\cN$.  We now examine the two separate cases. 
 
\subsection{$\cN$ non-null}   
 
Using $g^{ab}=h^{ab} + \epsilon n^a n^b$ reduces Eqn (\ref{ehboundary})  to 
\begin{equation} 
\label{nonnullterm} 
v^an_a = - n^a h^{bc} \{ \dn_a \delta g_{bc} -\dn_b \delta g_{ac}\}.
\end{equation} 
Comparing with the  variation of the extrinsic curvature $K$ of $\cN$ we see that 
\begin{eqnarray} 
\label{extrinsic} 
-2 \delta K &=&  2 h^{ab} C^c_{ab} n_c - 2 h^{ab} \dn_a \delta n_b\\ \nonumber  
&=&   - n^a h^{bc}  \dn_a \delta g_{bc} + 2 n^a h^{bc} \dn_b \delta g_{ac} + 2 \alpha K. 
\end{eqnarray} 
The first {terms} in  Eqn (\ref{nonnullterm})  and Eqn (\ref{extrinsic}) are the same. In \cite{wald,poisson}
the second term in Eqn (\ref{nonnullterm})  and the remaining terms  in Eqn (\ref{extrinsic}) 
are  put to zero but this unnecessarily restricts the allowed variations. Allowing the full 4-parameter
variation the second term in  Eqn (\ref{nonnullterm}) reduces to 
\begin{equation} 
n^a h^{bc} \dn_b \delta g_{ac}  = -2 \alpha K - h^{ab} \dn_a t_b,      
\end{equation} 
so that 
\begin{equation}
v^an_a=-2 \delta K + h^{ab} \dn_a t_b. 
\end{equation} 
Thus, in agreement with the standard results in \cite{wald,poisson} 
\begin{equation} 
\label{nnullboundarycorner} 
-\delta I_B+ \delta I_K = \frac{1}{2} \int_\cN  d^3x \sqrt {\epsilon h} D_a t^a , 
\end{equation}   
where $D_a$ is the connection compatible with $h_{ab}$ and 
\begin{equation} 
I_K=\int_\cN \sqrt{\epsilon h}K.  
\end{equation}  
If  $\corner_i \subset \cN$ are either spacelike or 
timelike  ``corners'' of $\cN$ with normals
$m_{(i)}^a \in T\cN$,  the variation Eqn (\ref{nnullboundarycorner}) reduces to  
\begin{equation}
\label{nnullcorner} 
\frac{1}{2} \sum_i \int_{\corner_i}d^2x  \sqrt{\epsilon' q}\,\, t^a m_{{(i)}a}  \equiv \sum_i
\delta I_{\corner_i}
\end{equation} 
where $q_{ab}$ is the induced metric on $\corner_i$ and   $\epsilon'=\pm 1$ depending on whether $\corner_i$ is spacelike or timelike.  If $\corner_i$ is
null, then 
\begin{equation}
\label{nulljoint} 
\delta I_{\corner_i}=\frac{1}{2}\int_{\corner}dx d\lambda \sqrt{\tilde{q}} \, \,  t^a \tk_a  
\end{equation} 
 where $\sqrt{tilde{q}}$ is the volume element on the $1$ dimensional spatial section of $\corner$
 and $\tk^a$ its null normal.  \red{As we will see  in the next few sections, such corner terms will not
   contribute}. Thus, the boundary term to be added to the action is 
\begin{equation}
I_B=I_K -\sum_i I_{\corner_i} 
\end{equation} 
where $I_{\corner_i}$ are the yet to be determined corner terms. 
\subsection{$\cN$ null} 

Since the null geodesics generated by $n^a$ are hypersurface orthogonal, we can suppose that they satisfy the 
condition $\dn_{[a} n_{b]}=0 $. Combining this with the variation $\delta g_{ab}=2n_{(a}
g_{b)c}\delta l^c$ allows us to simplify Eqn (\ref{ehboundary}) to
\begin{equation}  
\label{nullterm} 
v^an_a = n^a \dn_a \alpha -\alpha \Theta + 2 \alpha \kappa,   
\end{equation} 
where $\Theta$ and  $\kappa$ are the null expansion and surface gravity of $\cN$ respectively.

The natural analog of the  GHY term is 
\begin{equation} 
I_{\Theta} = \int_\cN d^2x d \lambda \sqrt{\sigma} \Theta 
\end{equation} 
and it  is therefore natural to first compare this variation with
Eqn (\ref{nullterm}).   Since $n^a=(\partial/\partial \lambda)^a$ remains invariant under this
class of  variations the affine parameter
$\lambda$ is unchanged, so that $\delta I_\Theta$ again only involves the integrand $\Theta$. While $\delta
\sigma_{ab}=0$,  
\begin{equation}
 \delta \sigma^{ab}=\delta g^{ac} g^{bd} \sigma_{cd} + g^{ac} \delta g^{bd} \sigma_{cd} = 2n^{(a}s^{b)} 
\end{equation} 
where we have used Equations (\ref{pert}) (\ref{npert}) so that 
\begin{eqnarray} 
\label{theta} 
2 \delta \Theta &=&  4n^{(a}s^{b)}\dn_a n_b  -2\sigma^{ab}   C^c_{ab} n_c + 2 \sigma^{ab} \dn_a \delta n_b \nonumber  \\ 
&=& -4 \kappa s^bn_b   - 2 \alpha \Theta + 2 \alpha \Theta =0. 
\end{eqnarray} 
Given the form of Eqn (\ref{nullterm}) it is therefore clear that an additional boundary piece is
required. Instead, consider  (see \cite{Parattu:2015gga,Parattu:2016trq})
\begin{equation} 
I_\kappa=    \int_\cN d^2x d \lambda \sqrt{\sigma} \kappa,
\label{nnullbdry}  
\end{equation}  
whose variation again only involves the integrand $\kappa$, 
\begin{eqnarray}
\label{kappa}  
2 \delta \kappa &=&2 ( \delta l^a n^b \dn_b n_a - l^a n^b C_{ab}^c n_c + l^a n^b\dn_b \delta n_a)
\nonumber \\ 
&=& 2 \alpha \kappa + 2  n^b \dn_b \alpha. 
\end{eqnarray} 
Thus 
\begin{eqnarray} 
\label{nullleftover}
-\delta I_B &=& \delta I_\kappa -\frac{1}{2}\int_\cN d^2x d\lambda \sqrt{\sigma} 
(n^a \dn_a \alpha + \alpha \Theta) \nonumber \\ 
&=& \delta I_\kappa-  \frac{1}{2}\int_\cN d^2x d\lambda \frac{d(\sqrt{\sigma} \alpha)}{d\lambda} 
\nonumber \\  
&=&  \delta I_\kappa+\delta I_\corner(\lambda_i)- \delta I_\corner(\lambda_f) 
\end{eqnarray} 
where  we have defined 
\begin{equation}
\label{nullcorner} 
\delta I_\corner(\lambda) \equiv \frac{1}{2} \int_\corner d^2x \sqrt{\sigma(\lambda)}\alpha(\lambda) 
\end{equation} 
and have used the expression $\Theta=\frac{1}{\sqrt{\sigma}} \frac{d \sqrt{\sigma}}{d
  \lambda}$.  Here, $\lambda_{i,f}$ are the initial and final values of $\lambda$ 
at the spacelike boundaries, $\corner_i, \corner_f$ of $\cN$.  As one can see, it is {\it only} such
spacelike corner terms that contribute for
$\cN$ null; there is no contribution from a null corner. The boundary term to be added to the action  is therefore 
\begin{equation}
\label{nullbdry}
I_B=-I_\kappa+ I_\corner(\lambda_f)-I_\corner(\lambda_i), 
\end{equation} 
where $I_\corner(\lambda)$ is a yet to be determined null corner term contribution. At this point
$I_\Theta$ can also be included, though its variation vanishes. This brings the boundary term into
the same form as that obtained in the tetrad formulation.  

Before moving on to  a calculation of the corner terms it is worthwhile saying a little about the
question of uniqueness of the transverse vector $Q^a$.  
In the non-null case, it is easy to find a
unique transverse vector. For any  timelike or spacelike vector $r^a
\in T_pM$ we associate a  unique transverse subspace $R  \subset T_p\cM$ such that $r.v=0, \forall v^a
\in R$.  $r^a$ is then transverse to $\cN$  and if it is normalised to $\pm 1$ it is the unique unit normal
$n^a$. In the null case, the situation is a little more complicated since $l^an_a=-1$ does not give
a unique $l^a$ associated to every $n^a$.  
We can however enforce uniqueness as follows. If
$m_1^a,m_2^a$ are spacelike unit vectors in $T_p\cN$ such that $n.m_{1,2}=0$, let  
$M_1,M_2 $ be their  associated transverse subspaces, respectively.  Then $l^a \in M_1\cap N_1$ is the
unique transverse null vector satisfying $l^an_a=-1$.

\subsection{Corner Terms for Null-Null Boundary} 
\label{nullnullsec} 

\begin{figure}
  \centering
    {\includegraphics[scale=0.5]{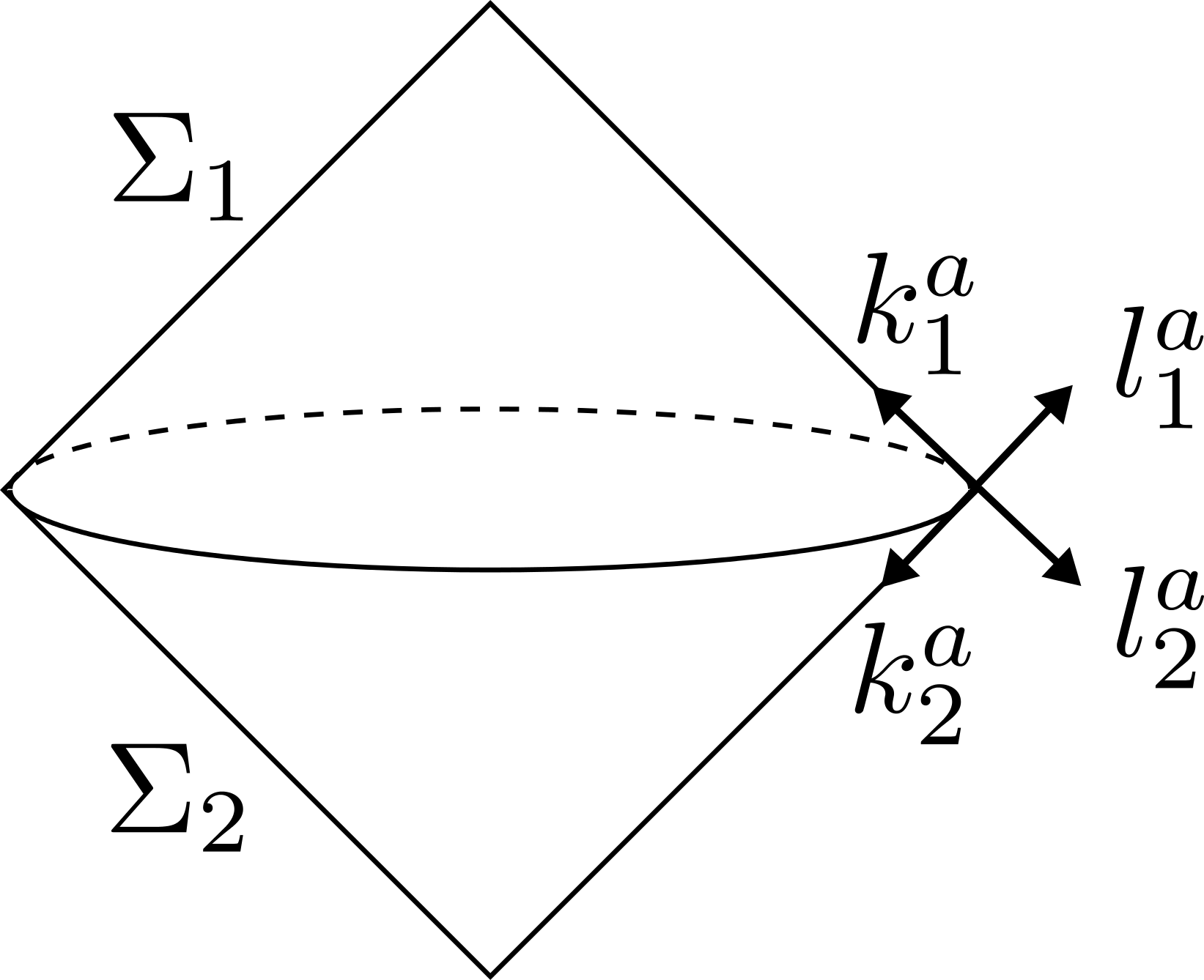}}
     \caption{An interval in $2+1$ Minkowski spacetime.}
     \label{fig:corner_terms}
\end{figure}

The intersection $\corner$ of two null hypersurfaces $\cN_{1,2}$ can be either spacelike or
null. Examples of these are shown in Figure~\ref{fig:corner_terms}. The achronality of a null hypersurface
precludes the intersection from being timelike.  When $\corner$  is null, as we have seen there
is no corner contribution for null $\cN$.  Indeed,  in any case, such a null intersection is not a join as
per our definition in Section \ref{two}. We therefore need only consider the case when
$\corner$ is spacelike.
For clarity in this section we will resort to using $k^a$ to depict the null normal and leave $n^a$
to denote the non-null normal. 

Let the normals to the two null boundary components $\cN_{1,2}$ be $k_{1,2}^a$.   In order to
fix the relative signs of the corner terms Eqn (\ref{nullcorner})  it is important to define first what is meant by an
outward pointing null normal. We will define this  using the transverse vector $l^a$ rather than the
normal $k^a$.   For a join arising in the causal diamond shown in  Figure~\ref{fig:corner_terms} the
 join is {\sl outward convex} in the following sense. The outward pointing transverse vector
 $l_1^a$ to $\cN_1$ in $\tcorner $ is along the positive $(\partial/\partial
v)^a $ direction. Hence $k_1^a$ is along the positive $(\partial/\partial u)^a $ direction. On the
other hand, $l_2^a$ for $\cN_2$ is in the negative $(\partial/\partial u)^a $ direction, which
makes $k_2^a$ lie in the negative $(\partial/\partial v)^a $ direction.  Thus, in this case the parameters
$\lambda_{1,2}$ on $\cN_{1,2}$ both take their initial values on $\corner$.  This is an outward
convex join. Conversely at an {\sl outward concave} join $\lambda_{1,2}$ both
take their final values on $\corner$. 

Thus, from Eqns (\ref{nullcorner}) and (\ref{nullbdry}) the total contribution to $\corner$ is 
\begin{equation} 
\delta I_\corner = \pm\frac{1}{2} \int_\corner d^2x \sqrt{\sigma}(\alpha_1 +\alpha_2)
\end{equation} 
depending on whether the join is concave or convex. Here $\alpha_{1,2}$ come from the variations of
$l_1^a$ and $l_2^a$ on $\corner$.  The variations of the metric on $\cN_1$ and $\cN_2$ are,
respectively
\begin{equation} 
\delta g^{ab}_{1,2} = k_{1,2}^{(a}( \alpha_{1,2} l_{1,2}^{b)} + \beta_{1,2} k^{b)}_{1,2}+s_{1,2}^{b)})
\end{equation} 
which at  $\corner$ must match up, i.e., 
\begin{equation} 
\label{jointeq}
\delta g^{ab}_1|_\corner=\delta g^{ab}_2|_\corner.      
\end{equation} 
Of the 4 null vectors $k_{1,2}^a, l_{1,2}^a$, we pick two linearly independent ones to be the
normals $k_{1,2}^a$, so that $ l_{1,2}^a= u_{1,2}k_1^a+v_{1,2}k_2^a$.  Using
$l_1.l_1=l_{2}. l_{2}=0$, $l_{1,2}.k_{1,2}=-1$, $v_1=u_2= -\frac{1}{k_1.k_2} $ and $u_1=v_2=0$,
which means that $l_{1,2}^a=-\frac{1}{k_1.k_2} k_{2,1}^a$.   Denoting the
equality on $\corner$ by $\doteq$, Eqn (\ref{jointeq}) then implies that
\begin{eqnarray} 
\label{equalities} 
\alpha_1  &\doteq& \alpha_2  \nonumber \\
\beta_1 &\doteq& \beta_2 \nonumber  \\ 
s_1^a \doteq s_2^a&\doteq&0   
\end{eqnarray} 
where we have used the linear independence of $k_{1}^a$ and $k_2^a$ and the directions tangent to $\corner$.
Noting that $\delta(k_1.k_{2}) = k_1. \delta {k_2} =- \alpha_2 (k_1.k_{2})$ allows us to express $\alpha_1$ as the variation 
$\alpha_1=-\delta(\ln(|k_1.k_2|)$ so that the corner term can be written as 
\begin{equation} 
\label{cornernullnull} 
I_\corner =  \mp \int d^2x \sqrt{\sigma} \ln(|k_1.k_2|), 
\end{equation} 
with the sign depending on whether $\corner$ is concave  or convex outward.

\subsection{Corner Terms for Null-Spacelike or Null-Timelike  Boundary}

The null-spacelike join can only be spacelike, while the null-timelike join can be either
spacelike or null. We will first consider the case when $\corner$ is spacelike.  
If  $\cN_1$ is  null and $\cN_2$ non-null, the corner contributions to the spacelike join
$\corner$ come from Eqn (\ref{nnullcorner}) and (\ref{nullcorner}) so that 
\begin{equation} 
\delta I_\corner= \pm \frac{1}{2}  \int_\corner d^2x \sqrt{\sigma} \alpha_1 -\frac{1}{2}  \int_\corner d^2x \sqrt{\sigma}  t^am_a, 
\end{equation}    
where $m^a$ is normal to $\corner$ in $\cN_2$ and the $\pm$ sign in front of the first term is
positive or negative if it is an initial or final boundary, respectively,  with respect to the
outward directed normal to $\cN_1$.  Again, we will see that $\corner$ can be thought of as concave 
or convex outward, and this determines the sign of the first term, but also of the second term. 

The variations of the metric on $\cN_1$ and $ \cN_2$
are  
\begin{eqnarray} 
\delta g^{ab}_1&=& k^{(a}( \alpha_1 l^{b)} + \beta_1 k^{b)}+s_1^{b)}) \nonumber \\
\delta g^{ab}_2 &=& n^{(a}(\alpha_2 n^{b)} + t^{b)}).  
\end{eqnarray} 
Decomposing $t^a \in \tsig_2$ $t^a=r^a+ s_2^a$, where $s_2^a\in \tcorner$ and $r^a$ is transverse to
$\tcorner$, and using the $k^a, l^a$ basis we express $n^a,r^a$ ( both transverse to $\tcorner$) as
$n^a=u_1 k^a + v_1 l^a, \quad r_2^a=u_2 k^a + v_2 l^a, $ where the normalisation $n^an_a=\epsilon
\Rightarrow v_1=\epsilon\frac{1}{2u_1}$.  Using the matching condition Eqn (\ref{jointeq}) and the
fact that $v_1$ can be arbitrary, we find that
\begin{eqnarray} 
\alpha_2&\doteq& -\frac{v_2}{v_1} \doteq -2 \epsilon u_1 v_2   \nonumber \\ 
\alpha_1 &\doteq& v_1(\alpha_2 u_1 + u_2) \nonumber \\ 
\beta_1& \doteq& u_1 (\alpha_2 u_1+u_2) \doteq2 \epsilon u_1^2 \alpha_1 \nonumber \\ 
s_1^a\doteq s_2^a&\doteq& 0.
\end{eqnarray}

\begin{figure}
    \centering
    \begin{subfigure}[b]{0.4\textwidth}
    \centering
        \includegraphics[width=0.7\textwidth]{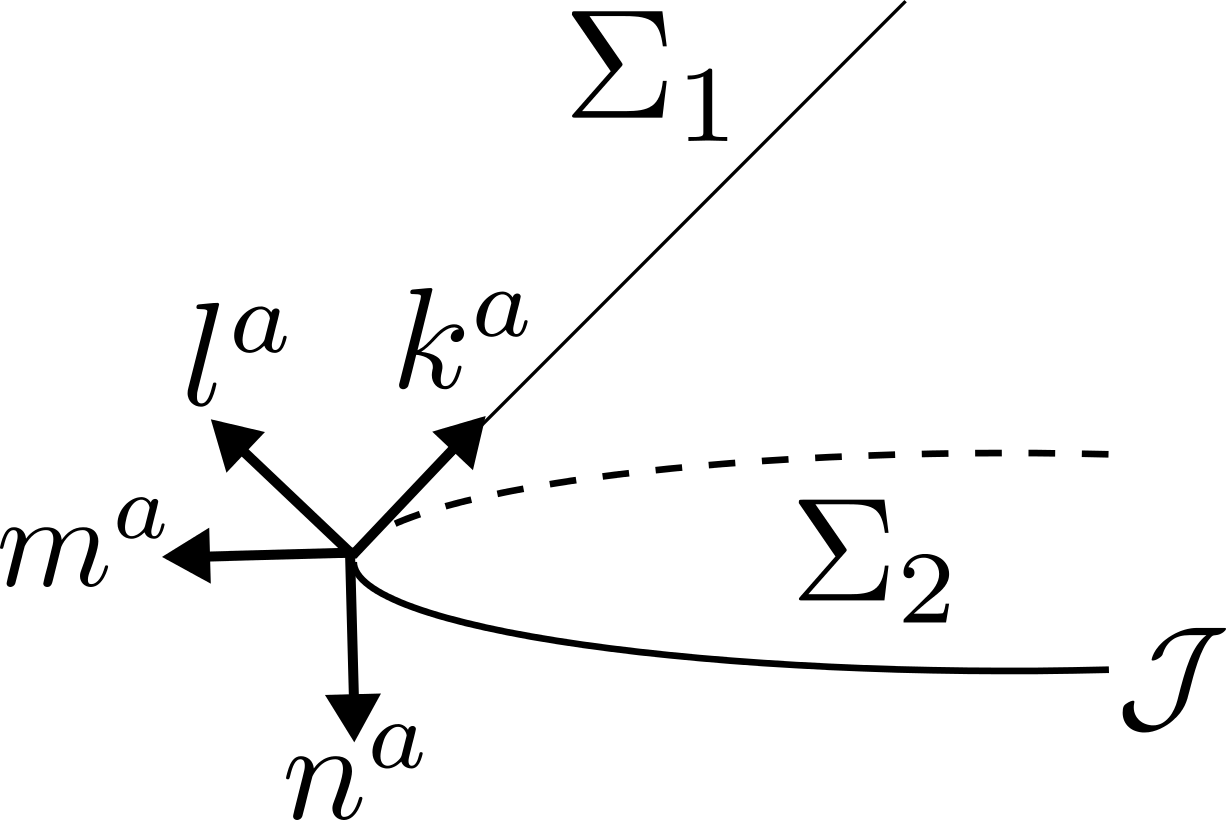}
        \caption{Future null $\Sigma_1$. Past spacelike $\Sigma_2$.}
        \label{fig:null_spacelike_1}
    \end{subfigure}
    \begin{subfigure}[b]{0.4\textwidth}
    \centering
        \includegraphics[width=0.7\textwidth]{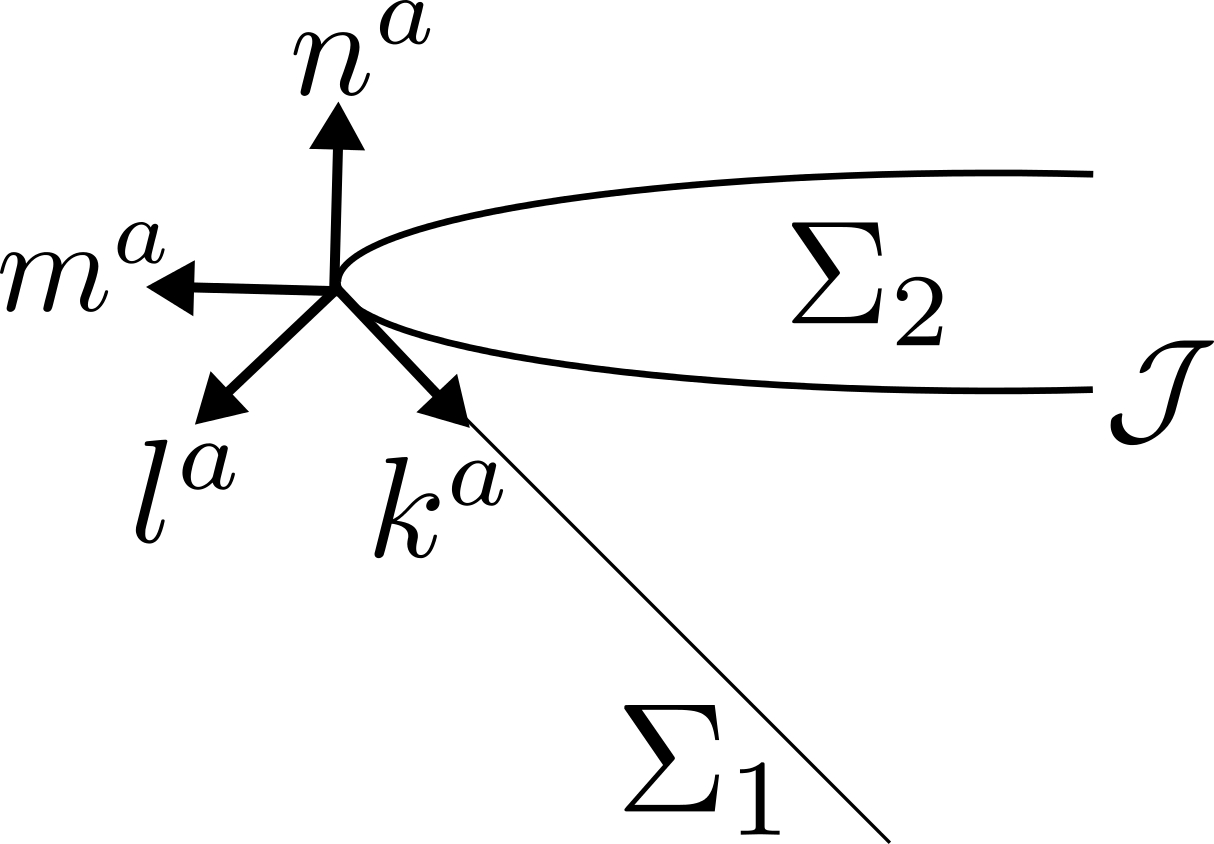}
        \caption{Past null $\Sigma_1$. Future spacelike $\Sigma_2$.}
        \label{fig:null_spacelike_2}
    \end{subfigure}

    \begin{subfigure}[b]{0.4\textwidth}
    \centering
        \includegraphics[width=0.6\textwidth]{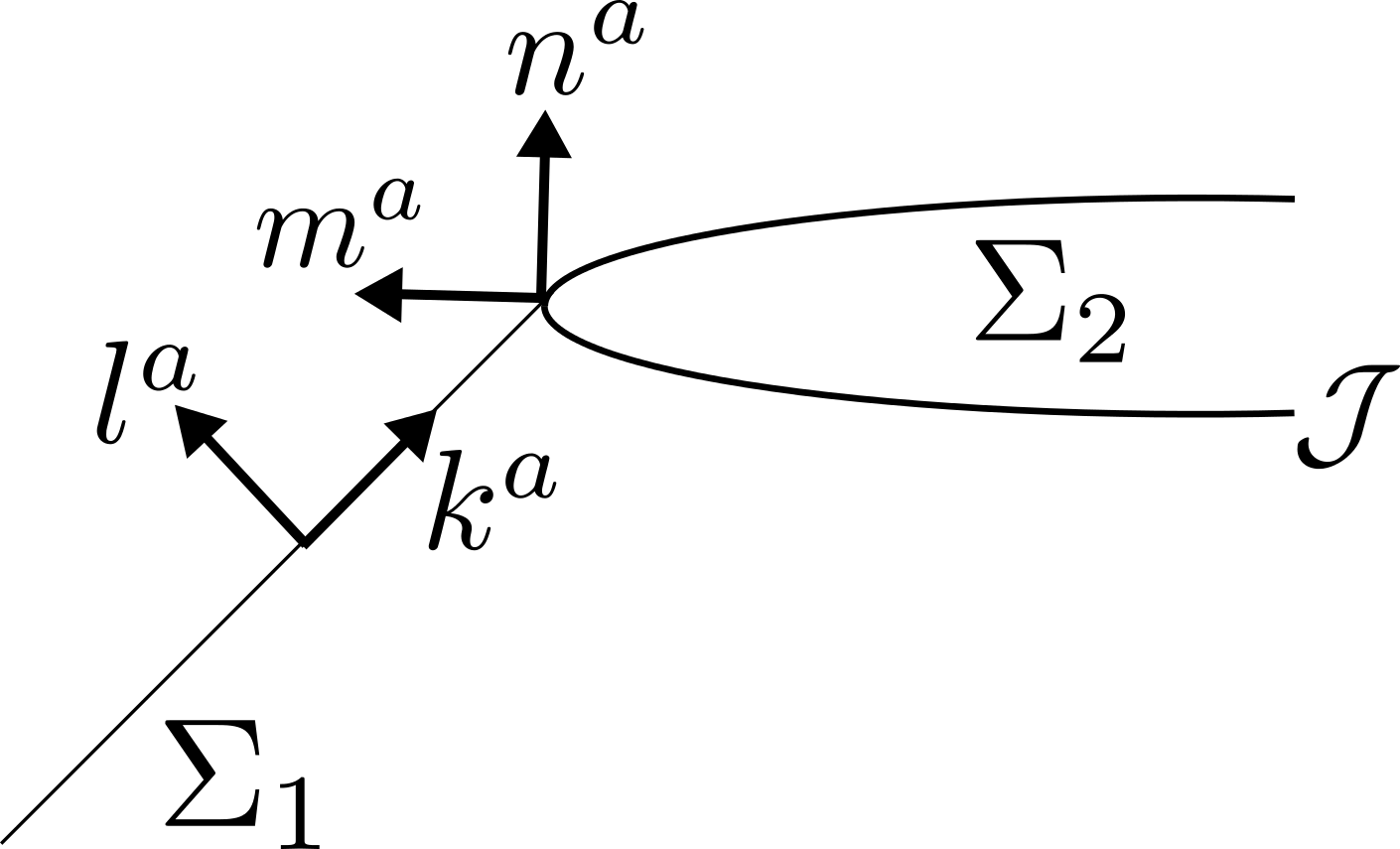}
        \caption{Future null $\Sigma_1$. Future spacelike $\Sigma_2$.}
        \label{fig:null_spacelike_3}
    \end{subfigure}
    \begin{subfigure}[b]{0.4\textwidth}
    \centering
        \includegraphics[width=0.7\textwidth]{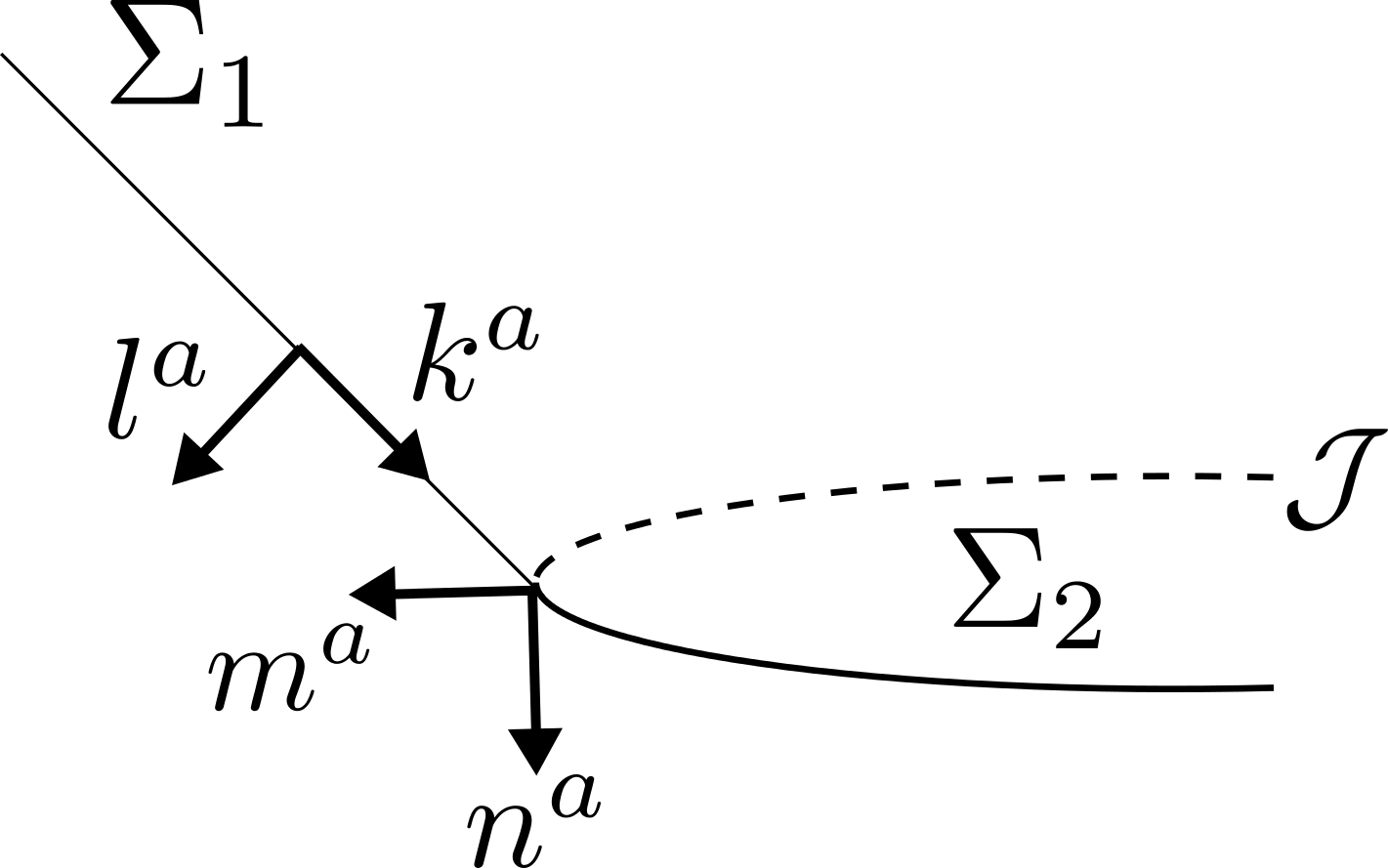}
        \caption{Past null $\Sigma_1$. Past spacelike $\Sigma_2$.}
        \label{fig:null_spacelike_4}
    \end{subfigure}
    
    \caption{Examples of null-spacelike joins in $2+1$ Minkowski spacetime showing the orientation of the vectors $k^a$, $l^a$, $n^a$ and $m^a$. The subcaptions illustrate whether a given null or spacelike surface is part of the future or past boundary of~$\mathcal{M}$.}\label{fig:null_spacelike}
\end{figure}

\begin{figure}
    \centering
    \begin{subfigure}[b]{0.4\textwidth}
    \centering
        \includegraphics[width=0.7\textwidth]{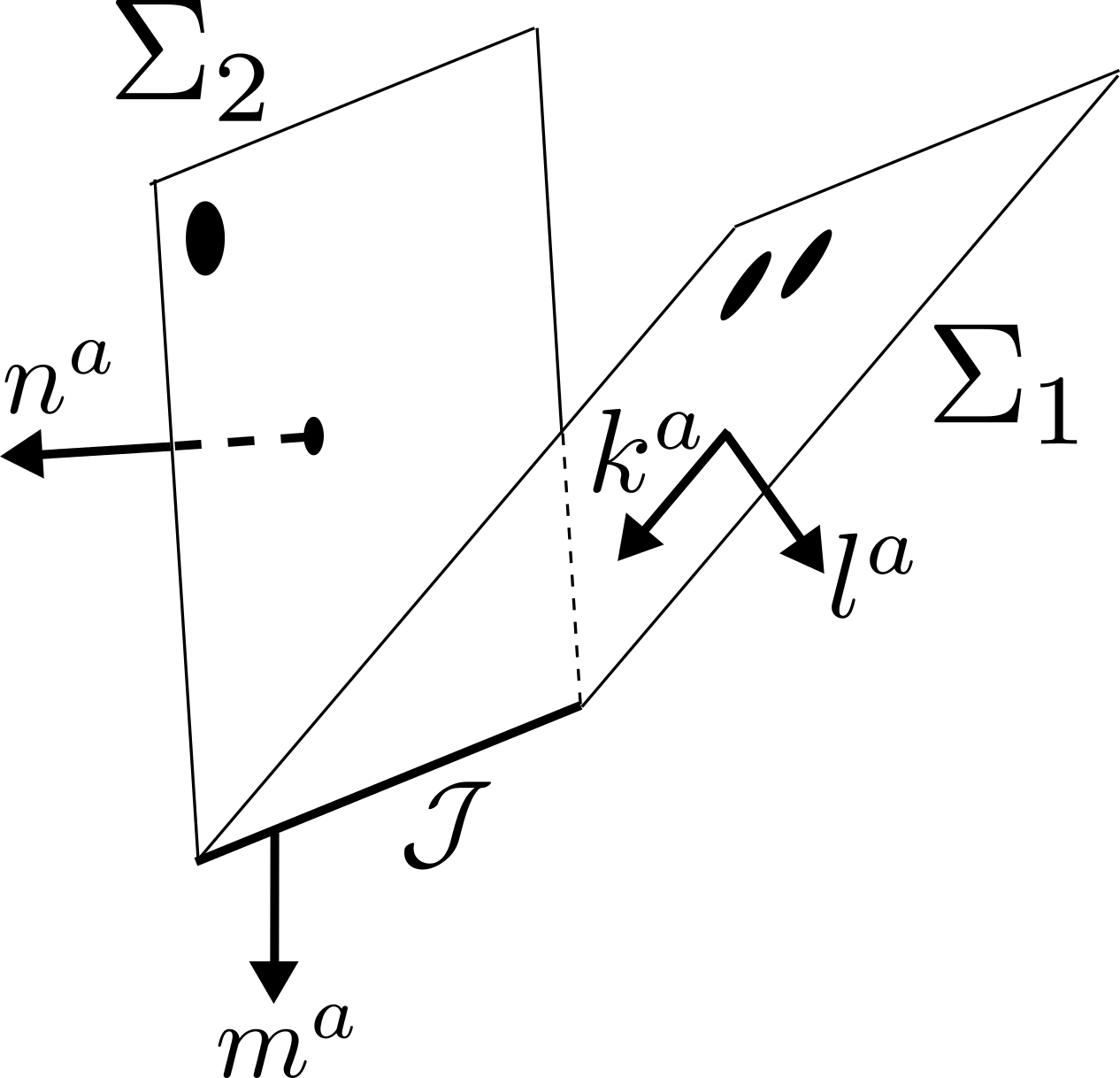}
        \caption{Past $\Sigma_1$ and $\mathcal{J}$.}
        \label{fig:null_timelike_with_joint_spacelike_1}
    \end{subfigure}
    \begin{subfigure}[b]{0.4\textwidth}
    \centering
        \includegraphics[width=0.7\textwidth]{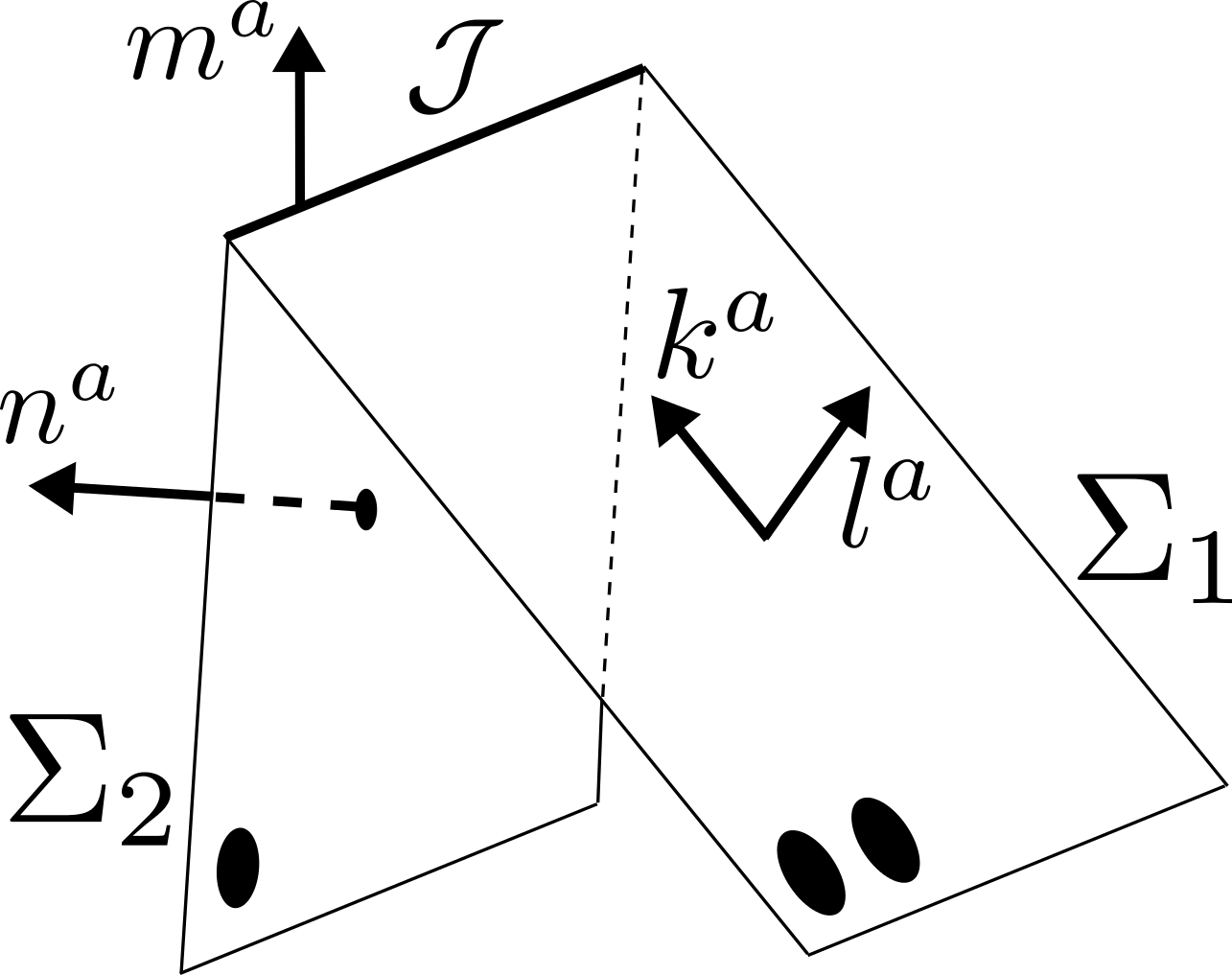}
        \caption{Future $\Sigma_1$ and $\mathcal{J}$.}
        \label{fig:null_timelike_with_joint_spacelike_2}
    \end{subfigure}

    \begin{subfigure}[b]{0.4\textwidth}
    \centering
        \includegraphics[width=0.6\textwidth]{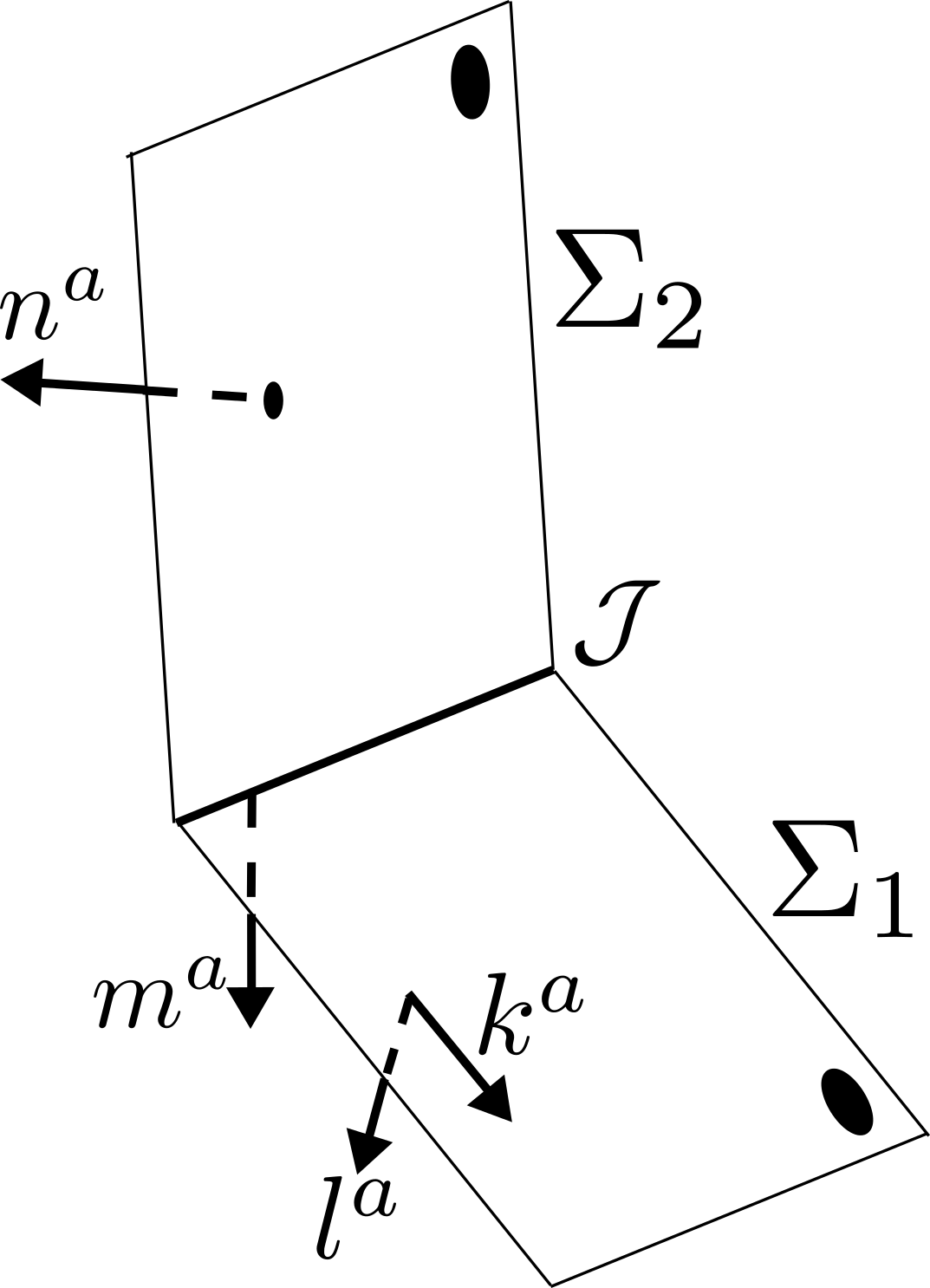}
        \caption{Past $\Sigma_1$ and $\mathcal{J}$.}
        \label{fig:null_timelike_with_joint_spacelike_3}
    \end{subfigure}
    \begin{subfigure}[b]{0.4\textwidth}
    \centering
        \includegraphics[width=0.7\textwidth]{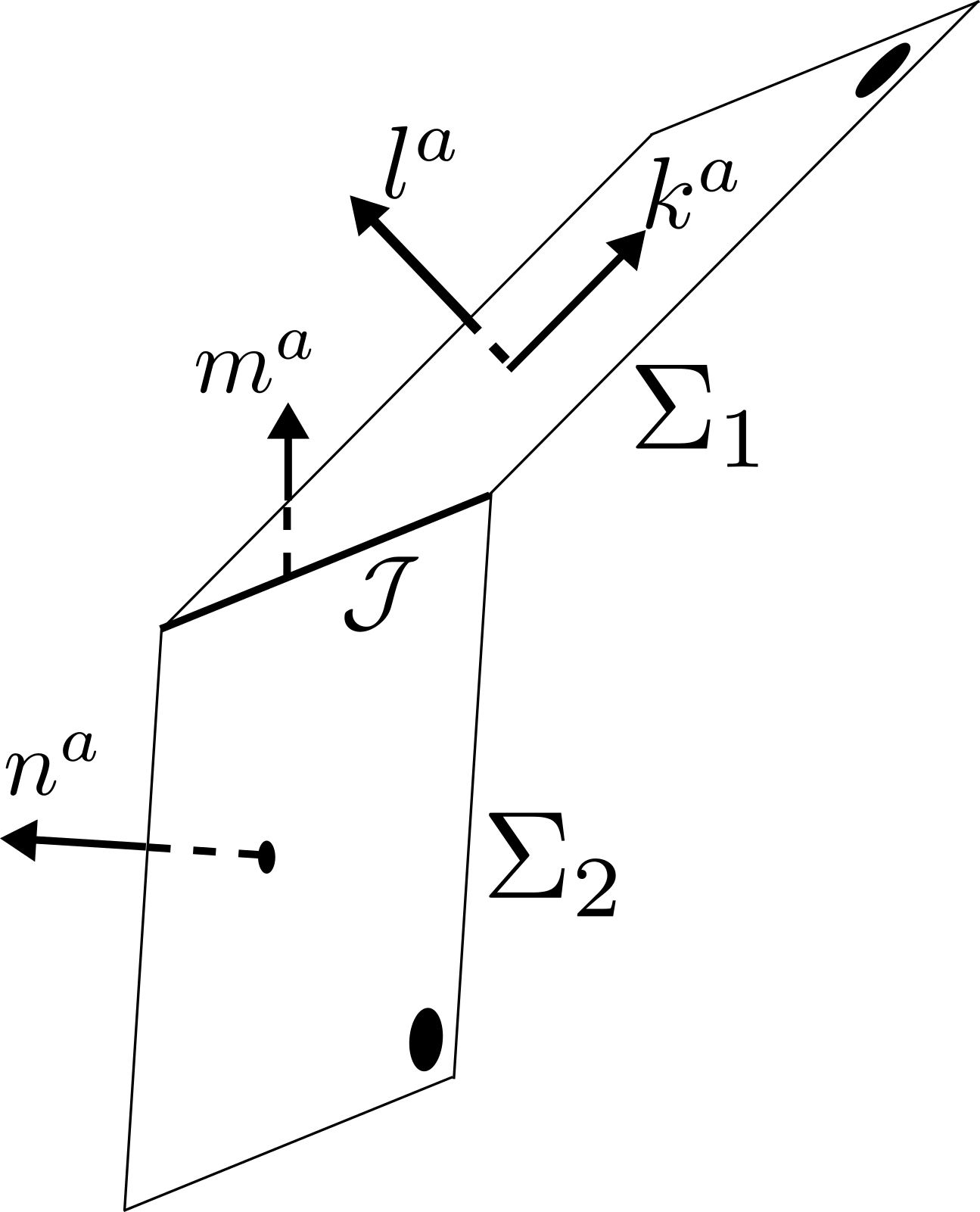}
        \caption{Future $\Sigma_1$ and $\mathcal{J}$.}
        \label{fig:null_timelike_with_joint_spacelike_4}
    \end{subfigure}
    
    \caption{Examples of the null-timelike case with a spacelike join in $2+1$ Minkowski spacetime. In each example we show a portion of the null and 
timelike surfaces $\Sigma_1$ and $\Sigma_2$ respectively. One (Two) dot(s) on a surface indicates that, from the perspective of the diagram, you are seeing the inside (outside) face of the surface with respect the region $\mathcal{M}$ that it bounds. The subcaptions illustrate whether the null surface and join are part of the future or past boundary of~$\mathcal{M}$.}\label{fig:null_timelike_with_joint_spacelike}
\end{figure}

Since the normal to a spacelike $\corner $ in $\tsig_2$ is spacelike when  $\cN_2$ is spacelike
and  timelike when $\cN_2$ is timelike, $m^am_a=-\epsilon$. Combining this with $n^am_a=0$,  we can
express  $m^a=\epsilon|u_1|k^a +\frac{1}{2|u_1|}l^a$.  Here we use the fact that since $m^a$ is outward
directed with respect to $\cN_2$,  it is $\epsilon$ times the  sense of the outward directed
$k^a$  as shown in Figures~\ref{fig:null_spacelike} and~\ref{fig:null_timelike_with_joint_spacelike}.

Thus, 
\begin{equation}
t^am_a=\frac{\epsilon}{2|u_1|}(\alpha_2
u_1+u_2)= \alpha_1 \frac{|u_1|}{u_1},  
\end{equation} 
where $\frac{|u_1|}{u_1} =\pm 1$ depending  on the  orientation of $\cN_2$ with
respect to $\cN_1$.  Specifically, $u_1=n.l$,  the  inner product of the transverse vectors
(which determine the ``outward'' directions )  of
$\cN_1$ and $\cN_2$. When $u_1<0$,  $\corner$ is an initial boundary with respect to the affine
parameter $\lambda$ on $\cN_1$, and when $u_1>0$, $\corner$ is a final boundary. Thus, 
\begin{equation} 
\delta I_\corner= \pm   \int_\corner d^2x \sqrt{\sigma} \alpha_1.
\end{equation} 

Again, using $\delta(n.k)= -\alpha_1 (n.k) \Rightarrow \alpha_1=-\delta(\ln(|n.k|))$ we find that
the corner term is 
\begin{equation} 
\label{cornernullnnull} 
I_\corner = \mp   \int d^2x \sqrt{\sigma} \ln(|n.k|). 
\end{equation} 

\begin{figure}
    \centering
    \begin{subfigure}[b]{0.4\textwidth}
    \centering
        \includegraphics[width=0.8\textwidth]{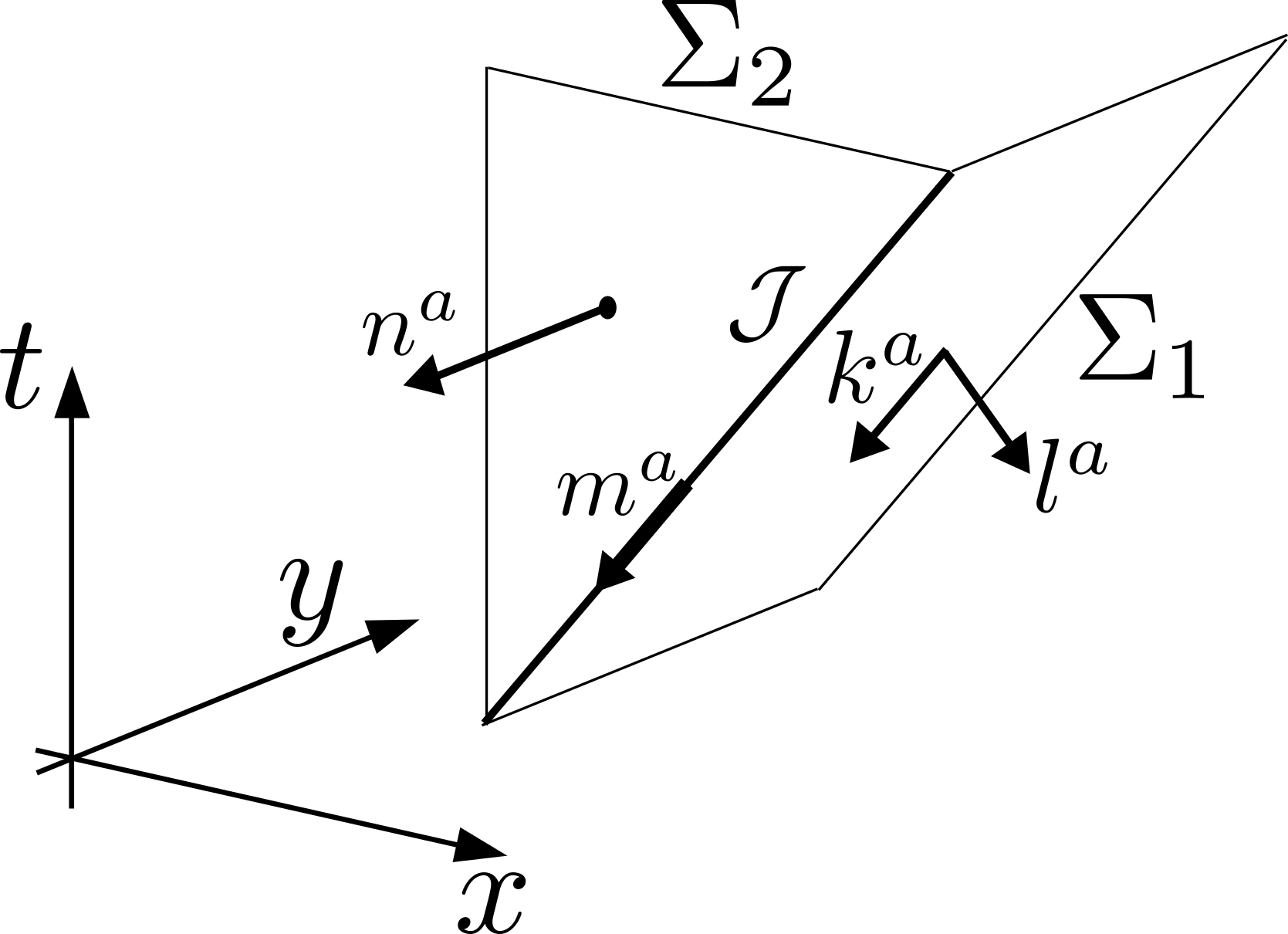}
        \caption{Past $\Sigma_1$ and $\mathcal{J}$.}
        \label{fig:null_timelike_with_joint_null_1}
    \end{subfigure}
    \begin{subfigure}[b]{0.4\textwidth}
    \centering
        \includegraphics[width=0.8\textwidth]{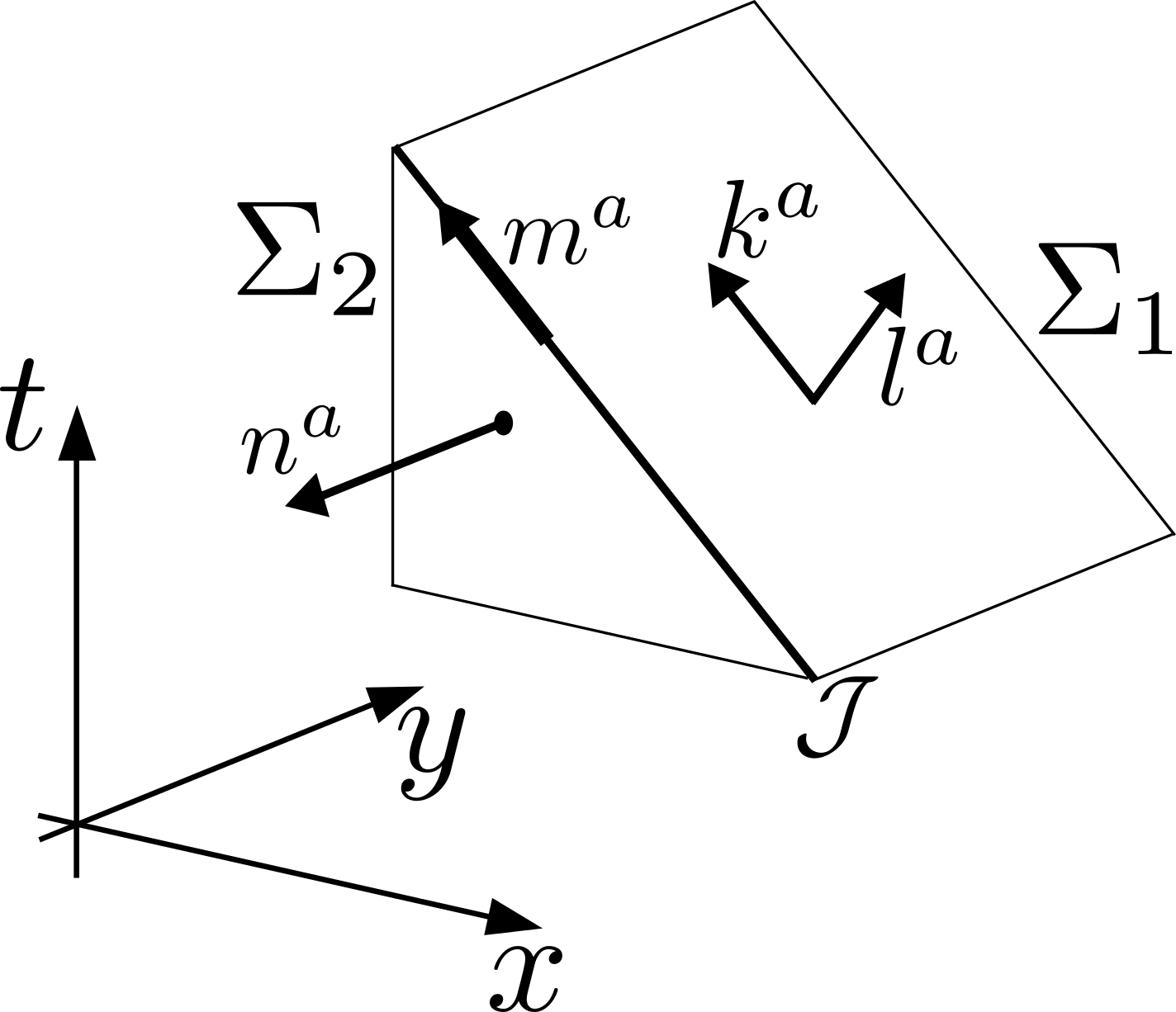}
        \caption{Future $\Sigma_1$ and $\mathcal{J}$.}
        \label{fig:null_timelike_with_joint_null_2}
    \end{subfigure}
    
    \caption{Examples of the null-timelike case with a null join in $2+1$ Minkowski spacetime. In each example we show a portion of the null and timelike surfaces $\Sigma_1$ and $\Sigma_2$ respectively. From the perspective of the diagram the outside faces of the surfaces can be seen, with respect the region $\mathcal{M}$ that they bound. The subcaptions illustrate whether the null surface and join are part of the future or past boundary of~$\mathcal{M}$.}\label{fig:null_timelike_with_joint_null}
\end{figure}

Finally,  let us consider the case when $\cN_2$ is timelike and $\corner$ is null. An example of
this is shown in Figure~\ref{fig:null_timelike_with_joint_null}. 

As discussed in Section \ref{nullnullsec} there is no contribution to  a non-spatial
corner from $\cN_1$, and the contribution from $\cN_2$ is given by
Eqn (\ref{nulljoint}).  Moreover, the normal $\tk^a$ to
$\corner$ coincides with that of $\cN_1$, i.e., $\tk^a=k^a$. Choosing the spatial basis
$\{\ts^a,\hs^a \}$ on $\cN_1$ such that $\ts^a$ is in $\tcorner$ and noting that $n^ak_a=0$,   $n^a
= w_1\hs^a$,  with  $n.n=1 \Rightarrow w_1^2=1$.  If we expand $t^a= u_2k^a+v_2 l^a+w_2\hs^a +z_2
\ts^a$, $n^at_a=0 \Rightarrow w_2=0$.  
 The variations  of the metric 
\begin{eqnarray} 
\delta g^{ab}_1&=& k^{(a}( \alpha_1 l^{b)} + \beta_1 k^{b)}+ s_1^{b)}) \nonumber \\
\delta g^{ab}_2 &=& n^{(a}(\alpha_2 n^{b)} + t^{b)}).  
\end{eqnarray} 
Expanding  $s_1^a=\gamma_1 \hs^a+ \gamma_2 \ts^a$,  
the matching condition Eqn (\ref{jointeq}) implies  that all the variables in the variation except
for $u_2$ and $ \gamma_1$ which are related by $w_1u_2=\gamma_1$, vanish  on $\corner$. 
Since $t^a\tk_a=-v_2= 0$ there is no corner term contribution. This is consistent with the fact that
the inner product of the normals $k.n=0$ and that $\delta(n.k)=\alpha_2(k.n)=0$.      
\red{
\subsection{Reparametrisation and the Null Boundary Action} 
\newcommand{\ttk}{\tilde k}
\newcommand{\tkap}{\tilde {\kappa}}
\newcommand{\tlam}{\tilde \lambda}
Combining all the boundary terms we find that 
\begin{equation} 
I_B= \sum_i I_{K_i} -\sum_j I_{\kappa_j} + \sum_k I_{\corner_k}, 
\end{equation} 
where $i,j,k$ range over the number of non-null boundary components, the number of null boundary
components and the number of corners, respectively. The null boundary term Eqn (\ref{nnullbdry})
is not invariant under reparametrisation. Let us consider the reparametrisation of the null vector 
\begin{equation}
{\ttk}^a=f(\lambda, x) k^a.   
\end{equation} 
where $f$ is strictly positive and $x$ is a local coordinate on the null generators.
The surface gravity associated with $\ttk^a$ then transforms as 
\begin{equation} 
\tkap= f(\lambda, x) \kappa - \frac{df}{d\lambda},  
\label{kappatrn}
\end{equation} 
so that 
\begin{eqnarray} 
 I_{\tkap} &=&  \int_\cN d^2x  \sqrt{\sigma} (d \tlam 
\tkap) \nonumber \\ 
&=&  \int_\cN d^2x  \sqrt{\sigma} (d \lambda \kappa) -    \int_\cN
d^2x  \sqrt{\sigma} (\frac{d \ln f(\lambda,x)}{d\lambda}) \nonumber \\ 
&=& I_\kappa -   \int_{\corner_f}  d^2x  \sqrt{\sigma} \ln f(\lambda_f,x)
+ \int_{\corner_i}  d^2x  \sqrt{\sigma} \ln f(\lambda_i,x)+\int_\cN d^2x  
 d \lambda \frac{d\sqrt{\sigma}}{d \lambda}[\ln f(\lambda,x)] .  
\label{array}
\end{eqnarray} 
The second and third terms exactly cancel the corner contribution (Equations(\ref{cornernullnull}), (\ref{cornernullnnull}))
\begin{equation} 
\mp   \int_\corner d^2x  \sqrt{\sigma} \ln (|\ttk.n|) =  \mp \int_\corner d^2x  \sqrt{\sigma} \ln
(|k.n|) \mp  \int_\corner d^2x  \sqrt{\sigma} \ln f(\lambda,x), 
\end{equation} 
which is negative or positive depending on whether $\lambda|_\corner$ is an final or initial value.    
Here $n^a$ represents the  normal to the ``other'' surface at the join $\corner$ which can be
either null or non-null.   
The presence of the last term in Eqn (\ref{array}),
which can be rewritten as
\begin{equation} 
\Delta I_B= \int_\cN d^2x\sqrt{\sigma}   \Theta(\lambda,x) d \lambda \ln f(\lambda,x) 
\end{equation} 
shows that the boundary action is {\it not} invariant under reparametrisation.
Let us now interpret this. Let us note first that under allowed variations,
(those that hold the boundary geometry fixed) the {\it variation}
of $\Delta I_B$ vanishes, since it depends only on the boundary geometry.
Thus the variation of the boundary action is reparametrisation invariant
although the action itself is not. As a general rule, it is differences in the action that
are important. Presumably, we can assume this to be true in quantum graity as well.

Recall the discussion in section 3.1, where we noted that the boundary
action is not gauge invariant under general gauge transformations  
although its {\it variation} is. This is exactly what is happening here. 
The surface gravity $\kappa$ is a component of a connection and (as seen in
Eqn (\ref{kappatrn})) transforms inhomogeneously under gauge transformations.
Reparametrisation changes the ``size'' of the null normal $k$ and is therefore
not in the little group.  The behaviour of the boundary action under reparametrisation 
is an example of the general phenomenon discussed there.

We also clarify that this lack of reparametrisation invariance of the null boundary action 
does not result in any arbritrariness in physical quantities. This is because 
what appears in physical quantities is the {\it difference} of two connections, which is a gauge covariant quantity.
This point is explained further in the conclusion. 

If one wishes, one could add ``counter terms'' to the boundary 
action to render it reparametrisation invariant. For example
\begin{equation}
-\int_\cN d^2x\sqrt{\sigma} d \lambda [\Theta(\lambda,x) \ln \frac{d\lambda}{dt} ],
\label{counterterm1}
\end{equation}
with $t$ being an arbitrary affine parameter, does the job. Another possibility\cite{LMPS} is 
\begin{equation}
-\int_\cN d^2x\sqrt{\sigma}  d \lambda[\Theta \ln \Theta]. 
\label{counterterm2}
\end{equation}
A third possibility is
\begin{equation}
-1/2 \int_\cN d^2x\sqrt{\sigma}  d \lambda[\Theta \ln{s_{ab}s^{ab}} ],
\label{counterterm3}
\end{equation}
where $s_{ab}$ is the shear tensor of the null geodesic congruence ruling the null surface.
Of these, the first Eqn (\ref{counterterm1}) depends on a choice of affine parametrisation,
which brings in some arbitrariness, since the parameter $t$ can be rescaled by $t\rightarrow c(x) t$,
where $c(x)$ depends on the null generator. A more serious problem is that including this counterterm spoils the additivity 
of the action for regions separated by a null boundary. For, the notion of an ``affine'' parameter in general will depend 
on which region we use to define the affine parameter. The two counter terms will therefore differ in value and therefore spoil 
the additivity of the action, which was one of our prime motivations. 

The second and third Eqn (\ref{counterterm2},\ref{counterterm3})
do not suffer from this ambiguity. However, they too have a problem: the counterterm is not differentiable if the expansion or shear vanishes.
Our view is that there no real need to add a counterterm at all since the lack
or reparametrisation invariance does not manifest itself in physical quantities.

}
\section{Conclusion}
\label{five} 
The main new advance of this paper is the realisation that the tetrad formulation of Einstein's theory permits a unified
approach to boundaries of all signatures. The calculations are considerably simplified and the use of differential forms permits us
to integrate over boundary manifolds regardless of their signature. Our derivation of the corner terms too is extremely simple. 
Our methods are complementary to \cite{Parattu:2015gga,Parattu:2016trq,LMPS} and our perpective is somewhat different. 
The differential form version of the boundary term
also makes it obvious that the boundary corrected action is additive. In any splitting of a spacetime into pieces, the boundary
term $I_B$ Eqn (\ref{boundaryterm2}) appears twice on the shared boundary with opposite orientation and so cancel out.  \red{The gauge non invariance
of the boundary action does not affect us here since the difference of the two connections is a gauge covariant object. In particular, the reparametrisation 
non invariance of the null boundary action does not spoil the additivity of the action.}

In this paper we have worked within the Dirichlet formalism for gravity in which 
the pullback metric $q_{ab}$ is held fixed on the boundary during the variation. One can also conceive of ``Neuman gravity'' 
in which the conjugate variable is held fixed. For example if the boundary is spacelike, the quantity 
$\sqrt{q}(K^{ab}-1/2 K q^{ab})$ related to the extrinsic curvature is conjugate to the three-metric.
There has been recent work \cite{chetan} 
exploring this possibility, albeit in the Euclidean context. 
Such alternate formalisms are of interest since it is far from clear which ensemble would 
prove the most advantageous in quantisation. It is also possible that these different choices may lead to different quantum theories.
For example, it is known in statistical mechanics that conjugate ensembles may not always be equivalent. Such issues are particularly acute
in the case of long range forces like gravity. A classic example is the stability question of a black hole in equilibrium with thermal radiation in a box.

A notable feature of the boundary term Eqn (\ref{boundaryterm2}) is that it is not gauge invariant
although its {\it variation} is. One must bear in mind that the boundary action is only determined
up to a functional of the boundary data that is held fixed, in our case the pullback of the metric
to the boundary.  One may worry that the value of the action changes under change of gauge. However,
there is no cause for concern. In a path integral formulation observable quantities are related to
the absolute value squared of the Feynman amplitude in Eqn (\ref{feynmanone}). This leads to a closed
time path integral of the Schwinger-Keldysh formalism. The quantity that appears in the exponent is
now $S(X_3,\Gamma)-S(X_3,\overline{\Gamma})$, where $\Gamma$ and $\overline{\Gamma}$ are histories
going between $X_1$ and $X_3$. While the two histories share the same final geometry $X_3$, they
have different values of the connection at the final point.  The two boundary terms at $X_3$ then
combine to give a gauge invariant answer, since the {\it difference} of two connections transforms
homogenously.  Another situation that arises is when one considers asymptotically flat spacetimes,
takes the boundary to infinity and interprets the boundary term in terms of the total mass. In this
case as is well known, we need to make a background subtraction in order to get a finite
answer. Once again, this subtraction results in a gauge invariant boundary term, since the
difference of two connections is a gauge covariant object. The gauge non invariance of the boundary
term is precisely what we have exploited in order to identify the corner terms.  This remark has a
parallel in the metric formulation too. The integrand in the boundary term Eqn (\ref{boundary}) is also
not coordinate invariant since it depends on the affine connection. The general allowed variation of
the metric Eqn (\ref{gpert}) can (at points of the boundary) 
be interpreted as a diffeomorphism generated by the vector field
$\xi_a=\phi Q_a$, where $\phi$ is any function that vanishes on the boundary. Under such a
diffeomorphism, the integrand in the boundary term changes by a total derivative and this permits us
to identify the corner terms in the metric formulation. 

\red{In the literature, it is suggested that the corner terms \cite{Neiman:2012fx} or their
  close analogs \cite{sorkinjorma} may pick up imaginary contributions.  (Imaginary
  contributions figure heavily in the Lorentzian Gauss-Bonnet theorem as
  well.)  Using our methods, such contributions would not be detected,
  as they have zero variation.}
However, the origin of such terms can be understood when the normal changes from timelike to spacelike.
We have chosen different adapted frames depending on whether the normal to the boundary is null, spacelike or timelike. This is because no Lorentz transformation
can connect these different normals. However, in connecting spacelike normals to timelike normals, it is possible to use complex Lorentz transformations.
If we complexify the Lorentz group to $O(2,\comp)$, the element $\Lambda=$
\[ \left( \begin{array}{ccc}
   \cosh{(\eta+i\pi/2)} & \sinh{(\eta+i\pi/2)}\\
   \sinh{(\eta+i\pi/2)}&\cosh{(\eta+i\pi/2)}\\
\end{array} \right)\] 
which has complex rapidity, $\eta+i\pi/2$ does the job of connecting spacelike and timelike normals. 
Thus every time the normal crosses  a null direction, (crossing counted with sign),
the action picks up an imaginary contribution $i\pi/2 \int dA$. This imaginary area term has been interpreted as  black hole entropy by Neiman and we refer the reader
to \cite{Neiman:2012fx} for a fuller discussion. 
While such a term affects the {\it value} of the Action, it does not affect the {\it variation},
since the variation of the area vanishes.
Note however, that no Lorentz transformation (real or complex) can relate a null normal to a spacelike or timelike one. It seems necessary to use different canonical
forms for null and non-null normals.

\red{The case of null boundaries has not receive much attention till the 
recent works of Neimann\cite{Neiman:2012fx,Neiman:2013ap,Neiman:2013taa,Neiman:2013lxa}, 
Parattu et al \cite{Parattu:2015gga,Parattu:2016trq} and Lehner et al \cite{LMPS}. Neimann was mainly 
interested in imaginary contributions to the action at the join of null boundaries.
He used affine parametrisations to describe the null generators, 
which is unnecessarily restrictive in the present context. 
The treatment of Parattu et al \cite{Parattu:2015gga} allows for arbitrary parametrisation
of the null generators and correctly identifies the form of the boundary 
action for null surfaces. However, these authors do not consider the corner terms,
which are necessary for a complete treatment of the boundary action. In a second 
paper \cite{Parattu:2016trq}, they attempt a unified description of both the null and non null case.
Their treatment is coordinate bound and makes assumptions about the behaviour of the normal
away from the boundary.
Lehner et al \cite{LMPS} provide a metric treatment of the null boundary terms and identify the corner terms.
They also have a detailed discussion of reparametrisation invariance and suggest counterterms to 
be added to the boundary action.

In the present work, we use the power of Cartan's tetrad formulation and 
differential forms to considerably simplify the treatment. Differential forms
give us a unified approach to boundaries of all signatures. We compute the corner terms quite
simply using the local Lorentz invariance of the tetrad formalism. In the mathematical section we also give a classification
of all possible corner signatures, including the case of null joins (see Figure 5) that have not been considered in the above works.
In order to reach a wider audience we also
translate our results into the metric language which is more familiar to readers. We have also noted the 
contribution which come from ``creases'' that appear in spacetimes with a dynamically evolving black hole 
exterior. Finally, we offer a perspective on reparametrisation invariance (RI) in the null case, which differs slightly
from Ref \cite{LMPS}. Rather than try to restore RI, we note that the lack of RI in the boundary action does not affect any physical quantity
in the path integral. 
}

We close with a remark regarding the asymptotics of gravitational fields.  Let us compare the
  value of the action in the second order Einstein-Hilbert form and the first order form. For
  asymptotically flat spacetimes, the metric tends to its flat asymptotic form $g_0$ at the rate
  $(g-g_0)=O(1/r)$. As a result, the difference between the connection $\Gamma$ of $g$ and the flat
  connection $\Gamma_0$, $\Delta \Gamma=\Gamma-\Gamma_0$ goes as $\Delta \Gamma = O(1/r^2)$ and $R=
  O(1/r^3)$. The Einstein Hilbert form diverges logarithmically at radial infinity ($\int R r^2
  dr\approx \int dr/r$) but the first order form converges: ($\int (\Delta \Gamma)^2 r^2
  dr\approx\int dr/r^2$).  This allows an interpretation of the 4-momentum as a well defined
  variation of the action, i.e as a Noether charge.  While there has been much work on null
  infinity\cite{bondi}, we are not aware of any discussion of boundary counterterms in this context,
  for example, in the derivation of the Bondi mass.  The issue of null boundaries has been neglected
  until the recent interest generated by \cite{Parattu:2015gga,Parattu:2016trq}. There has been
  recent work \cite{strominger,ashtekarnew} reviving the topic of asymptotic null infinity
  \cite{ashtekarone,ashtekartwo,ashtekarthree} and relating it to soft theorems in particle physics.
  We hope that our treatment of null boundaries may help understand null asymptotics of
  gravitational fields.

\section{Acknowledgements} 

This work was  supported in part under an agreement with Theiss Research and funded by a
grant from the FQXI Fund on the basis of proposal FQXi-RFP3-1346 
to the Foundational Questions Institute. 
I.J. is supported by the EPSRC. 
RDS's research was supported in part by NSERC through grant RGPIN-418709-2012.
This research was supported in part by Perimeter Institute for
Theoretical Physics. Research at Perimeter Institute is supported
by the Government of Canada through Industry Canada and by the
Province of Ontario through the Ministry of Economic Development
and Innovation.

We gratefully acknowledge 
an email correspondence with Yang Run Qiu, which led us to improve the paper.

\bibliographystyle{utphys}
\providecommand{\href}[2]{#2}\begingroup\raggedright
\endgroup
\providecommand{\href}[2]{#2}\begingroup\raggedright
\end{document}